\RequirePackage{pdf14}
\pdfoutput=1

\documentclass[10pt, conference]{IEEEtran}

\usepackage[a-1b]{pdfx}
\usepackage{cite}
\usepackage{listings}
\usepackage{multirow}
\usepackage{graphicx}
\usepackage{hyperref}
\usepackage{fancybox}
\usepackage{enumitem}
\usepackage{xspace}
\usepackage{tablefootnote}
\usepackage{booktabs}
\usepackage{subcaption}
\usepackage{xcolor}
\usepackage{caption}
\usepackage{algorithm}
\usepackage[noend]{algpseudocode}

\usepackage{float}
\newfloat{algorithm}{t}{lop}

\newcommand\takeaway[1]{
  \setlength{\fboxsep}{0.8em}
  \vspace{-1em}
  \begin{center}
  \Ovalbox{\begin{minipage}{0.95\linewidth}
    #1
    \end{minipage}}
  \end{center}
}
\newcommand\parentheticalstats[1]{{(\textit{#1})}}

\definecolor{lightgray}{gray}{0.9}

\begin{document}

\title{An Empirical Study of OSS-Fuzz Bugs}

\author{
	\IEEEauthorblockN{Zhen~Yu Ding\IEEEauthorrefmark{1} and Claire Le~Goues\IEEEauthorrefmark{2}}
	\IEEEauthorblockA{
		\IEEEauthorrefmark{1}Motional (work done at Carnegie Mellon University), Pittsburgh, USA \\
		zhd23@pitt.edu
	}
	\IEEEauthorblockA{
		\IEEEauthorrefmark{2}School of Computer Science, Carnegie Mellon University, Pittsburgh, USA \\
		clegoues@cs.cmu.edu
	}
}

\maketitle

\begin{abstract}
	Continuous fuzzing is an increasingly popular technique for automated
	quality and security assurance. Google maintains OSS-Fuzz: a continuous
	fuzzing service for open source software. We conduct the first empirical
	study of OSS-Fuzz, analyzing 23,907 bugs found in 316 projects.
	We examine the characteristics of fuzzer-found faults, the lifecycles of
	such faults, and the evolution of fuzzing campaigns over time.
	We find that OSS-Fuzz is often effective at quickly finding
	bugs, and developers are often quick to patch them. However, flaky bugs,
	timeouts, and out of memory errors are problematic, people rarely file
	CVEs for security vulnerabilities, and fuzzing campaigns
	often exhibit punctuated equilibria, where developers might be surprised
	by large spikes in bugs found. Our findings have implications on future
	fuzzing research and practice.
\end{abstract}


\section{Introduction}

Fuzz testing is effective at finding bugs and security vulnerabilities
such as crashes, memory violations, and undefined behavior.
Continuous fuzzing --- using fuzz tests as part of a continuous testing
strategy --- is increasingly popular in both
industry~\cite{clusterfuzz, mayhem, onefuzz, gitlab-fuzzing, code-intelligence}
and open-source software engineering~\cite{clusterfuzz, onefuzz, oss-fuzz, syzbot}.
To improve the quality and security of open-source, Google maintains
OSS-Fuzz~\cite{oss-fuzz}, a continuous fuzzing service supporting over 300
open-source projects.

We present the first empirical study of OSS-Fuzz, examining over 4 years of
data and 23,907 fuzzer-discovered bugs found in 316 software projects.
To our best knowledge, this is the largest study of continuous fuzzing
at the time of writing.
We expand the body of empirical research on fuzzing, which the fuzzing
community expressed a need for~\cite{shonan-fuzzing}.
Our main contributions are:
\begin{itemize}
	\item We present the first empirical study of OSS-Fuzz and the largest study of
	continuous fuzzing at the time of writing, analyzing 23,907 bugs in 316 projects.
	\item We analyze the characteristics of fuzzer-found bugs.
	We consider fault types, flakiness, fuzz blockers, unfixed bugs, CVE entries,
	and relationships among these features. We find that many fuzzer-found bugs
	harm availability without posing direct threats to confidentiality or integrity,
	timeouts and out of memory errors are unusually flaky, flaky bugs are mostly
	unfixed, and few bugs, mostly memory corruption bugs, receive CVE entries.
	\item We probe OSS-Fuzz bugs' lifecycles. We find that most fuzzer-found bugs are
	detected and fixed quickly, albeit lifecycles vary across fault types,
	flaky bugs are slower to detect and fix, and fuzz blockers are slower to fix.
	\item We study the longitudinal evolution of fuzzing campaigns.
	We find that bug discovery often show punctuated equilibria,
	with occasional spikes in the number of bugs found interspersed among relatively slow bug hunting.
\end{itemize}

Section~\ref{sec:background} provides background to contextualize our work.
Section~\ref{sec:oss-fuzz-bug-reports} overviews the bug reports under study.
We analyze fault characteristics, fault lifecycles, and longitudinal evolution
in Sections~\ref{sec:fault-characteristics}, \ref{sec:fault-lifecycle},
and \ref{sec:longitudinal-evolution} respectively. Section~\ref{sec:discussion}
discusses our findings' implications for research and practice,
Section~\ref{sec:related-work} overviews related work, and Section~\ref{sec:conclusion} concludes.

\section{Background}
\label{sec:background}

To explain how OSS-Fuzz found the bugs under analysis, we provide background
on coverage-guided fuzzing (Section~\ref{sec-fuzzing}) and OSS-Fuzz (Section~\ref{sec:oss-fuzz}).
We also discuss the CIA triad of information security (Section~\ref{sec:cia}), which we use to analyze
the security impact of bugs in our analysis.

\begin{algorithm}
\caption{Coverage-guided fuzzing.}
\label{alg:fuzzing}
\small{
\begin{algorithmic}[1]
	\Procedure{Fuzz}{program $p$, set of seed inputs $I_0$}
		\State \textit{Inputs} $\gets I_0$
		\State \textit{TotalCoverage} $\gets$ coverage of $p$ on \textit{Inputs}
		\While{within time budget}
			\State $i \gets$ pick from \textit{Inputs}
			\State $i' \gets$ mutate $i$
			\State \textit{coverage}, \textit{error} $\gets$ execute $p$ on $i'$
			\If{$\exists$ \textit{error}}
				\State report \textit{error} and faulty input $i'$
				\State optionally, exit
			\ElsIf{\textit{coverage} $\not\subseteq$ \textit{TotalCoverage}}
				\State add $i'$ to \textit{Inputs}
				\State \textit{TotalCoverage} $\gets$ \textit{TotalCoverage} $\cup$ \textit{coverage}
			\EndIf
		\EndWhile
	\EndProcedure
\end{algorithmic}
}
\end{algorithm}

\begin{figure*}
	\centering
	\includegraphics[width=\linewidth]{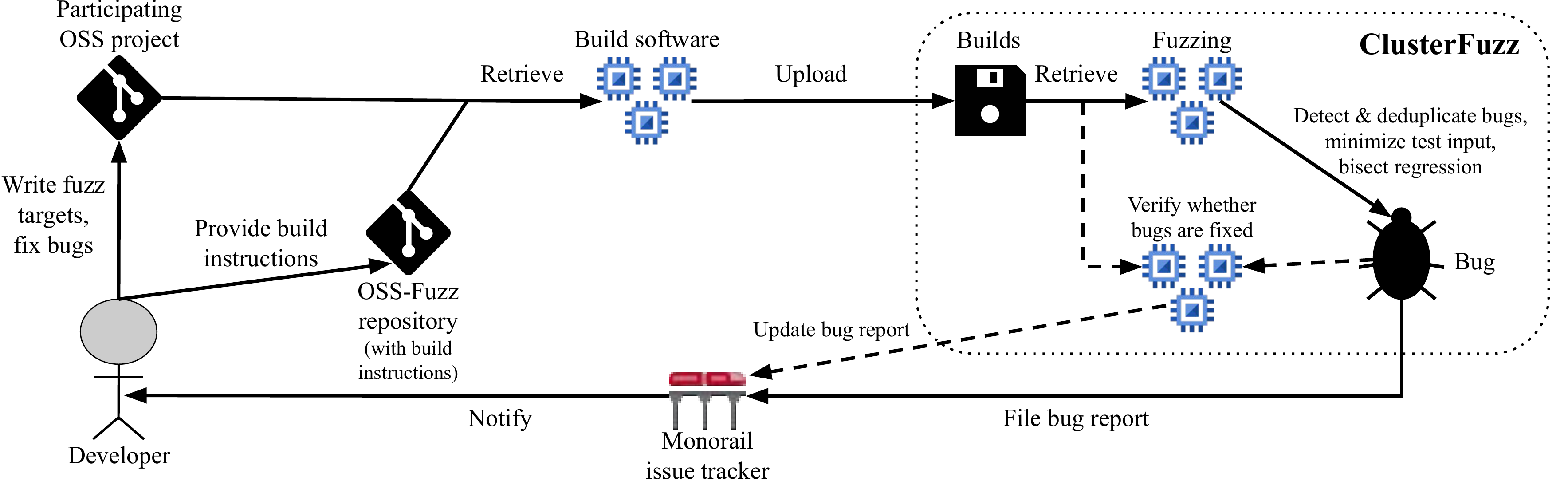}
	\caption{OSS-Fuzz's workflow}
	\label{fig:oss_fuzz_process}
\end{figure*}

\subsection{Coverage-guided fuzzing}
\label{sec-fuzzing}

Coverage-guided fuzzing (CGF), implemented by tools such as AFL~\cite{afl},
libFuzzer~\cite{libFuzzer}, and honggfuzz~\cite{honggfuzz}, is a popular bug detection method.
CGF uses genetic search to find inputs that maximize code coverage.
Algorithm~\ref{alg:fuzzing} describes CGF at a high level.
The algorithm maintains a pool of \textit{Inputs} and the \textit{TotalCoverage} of
program $p$ on \textit{Inputs}. The user provides seed inputs $I_0$ to instantiate
\textit{Inputs}. The fuzzer repeatedly picks an input $i$ from the pool of \textit{Inputs}
and applies a mutation (e.g., increment, bit flip, or user-defined mutations)
to produce $i'$. The fuzzer then executes program $p$ on mutated input $i'$
to gather the \textit{coverage} of program $p$ on $i'$ and detect any \textit{error},
such as crashes, assertion violations, timeouts,
memory leaks or access violations (with ASan~\cite{asan}), undefined behavior
(with UBSan~\cite{ubsan}), uninitialized memory use (with MSan~\cite{msan}),
or data races (with TSan~\cite{tsan}). If $i'$ does not trigger an $error$
and discovers new \textit{coverage} that is not previously seen in \textit{TotalCoverage},
then add $i'$ to \textit{Inputs} and update \textit{TotalCoverage}.
By finding inputs that cover new code, CGF aims to test as much of the
program as possible.

Fuzzers need an entrypoint into the program to provide test inputs;
such an entrypoint is often called a \emph{fuzz target}.
libFuzzer-style fuzz targets --- which AFL and honggfuzz also support
and OSS-Fuzz uses --- are functions that take in fuzzer-generated
arbitrary bytestream input, transform the input to program-usable input data
if needed, and execute the program under test with the input.

\subsection{Continuous Fuzzing and OSS-Fuzz}
\label{sec:oss-fuzz}

Continuous fuzzing uses fuzzing as part of a continuous testing strategy
to find regressions as software evolves. Several organizations
incorporate fuzzing as part of their quality assurance
strategy~\cite{clusterfuzz,onefuzz,dgraph-fuzz,military-fuzzing}
or offer tools that provide continuous fuzzing as a service~\cite{mayhem,gitlab-fuzzing,code-intelligence,oss-fuzz}.

OSS-Fuzz~\cite{oss-fuzz} is Google's continuous fuzzing service for open source
software (OSS) projects that are widely used or critical to global IT
infrastructure. OSS-Fuzz uses ClusterFuzz~\cite{clusterfuzz}, Google's
continuous fuzzing framework.
Figure~\ref{fig:oss_fuzz_process} illustrates OSS-Fuzz's workflow.
Developers in a participating OSS project write fuzz targets and provide instructions
for building the software. OSS-Fuzz continuously builds the software and uploads
it to ClusterFuzz. ClusterFuzz finds fuzz targets and uses the coverage-guided
fuzzers AFL~\cite{afl}, libFuzzer~\cite{libFuzzer}, and honggfuzz~\cite{honggfuzz} to
fuzz the software. Upon detecting a bug, ClusterFuzz checks whether the bug is a duplicate
of any previously found bugs, minimizes the bug-inducing input, and bisects the range
of commits in which the regression occurred. If the bug is not a duplicate, then ClusterFuzz
files a bug report on Monorail, an issue tracker. ClusterFuzz periodically verifies whether
any previously found bugs are fixed; if so, OSS-Fuzz updates fixed bugs' report.

Bug reports are initially available only to project members.
OSS-Fuzz uses Google's standard 90-day public disclosure
policy~\cite{oss-fuzz-bug-disclosure,projzero-bug-disclosure} for all found bugs.
If a bug is patched, then
the disclosure date moves up to either 30 days post-patch or stays at 90 days
post-discovery, whichever is earlier.
Bug disclosure deadlines are a standard practice in industry; deadlines encourage prompt
repair, while the delay in public disclosure gives developers time to write and discreetly
distribute patches.

\subsection{CIA Triad}
\label{sec:cia}

Since fuzzing is often used as part of a security testing strategy, we analyze
the potential security impacts of bugs using the CIA triad~\cite{ieee75-saltzer}.
CIA stands for confidentiality, integrity, and availability; each is
a desired security property.

Confidentiality entails that only authorized users can access a resource.
Data breaches are instances of violated confidentiality.
Dangerous memory reads, such as buffer overflow reads or use of uninitialized
values, can leak confidential data residing in memory.
For example, Heartbleed was a buffer overflow read vulnerability in OpenSSL
that jeopardized the confidentiality of server data, such as private keys~\cite{heartbleed}.

Integrity entails that only authorized users can modify resources in an allowable
manner. Unauthorized deletion or tampering of resources are instances of
violated integrity. Tampering of memory --- which may result from improper memory
management or dangerous functions that allow for buffer overflow writes or unsafe
heap operations --- can corrupt data or facilitate arbitrary code execution.

Availability entails that resources remain available to users.
Denial of service attacks harm availability.
Attackers can leverage resource exhaustion bugs such as timeouts, out of
memory errors, or memory leaks to reduce system performance. Bugs that
result in abnormal process termination such as null dereferences, stack overflows,
or operating system signals can deny service to anyone else using the same
terminated process.

\section{OSS-Fuzz Bug Reports}
\label{sec:oss-fuzz-bug-reports}

\begin{figure}
\footnotesize{
	\fcolorbox{lightgray}{lightgray}{
	\begin{minipage}{0.32\linewidth}
		\textbf{Issue 20000}

		Reported by ClusterFuzz

		on Fri, Jan 10, 2020

		7:15 AM EST
		\\

		Status: Verified (Closed)

		Modified: Feb 10, 2020
		\\

		Labels:

		Reproducible

		\textit{7 others\ldots}
	\end{minipage}}
	\begin{minipage}{0.68\linewidth}
		Project: binutils

		Fuzzing Engine: libFuzzer

		Fuzz Target: fuzz\_disassemble

		Platform Id: linux
		\\

		Crash Type: Unsigned-integer-overflow

		Sanitizer: undefined (UBSAN)

		Regressed: {\scriptsize{oss-fuzz.com/revisions?job=\textit{omitted}\\ \&range=201912170318:201912190318}}

		Reproducer Testcase: {\scriptsize{oss-fuzz.com/download\\ ?testcase\_id=\textit{omitted}}}
	\end{minipage}
	\begin{minipage}{\linewidth}
		\vspace{1.5mm}
		\textbf{Comment 1} by ClusterFuzz on Jan 11, 2020, 10:24 AM EST

		ClusterFuzz testcase\ldots is verified as fixed in \textit{link to fix code commit range.}

		\vspace{1.5mm}
		\textbf{Comment 2} by sheriffbot on Feb 10, 2020, 1:10 PM EST

		This bug has been fixed for 30 days. It has been opened to the public.
	\end{minipage}
	\caption{An OSS-Fuzz bug report. Some details are omitted or edited for brevity.
	Original report at \url{https://bit.ly/3oaLhCp}}
	\label{fig:bug-report}
}
\end{figure}

We extract data from OSS-Fuzz bug reports on Monorail.
Figure~\ref{fig:bug-report} shows OSS-Fuzz Issue \#20000 as an example.
ClusterFuzz generates these reports in a standardized format.
The report indicates which software ``Project'' is affected,
the ``Fuzzing Engine'' and ``Fuzz Target'' that found the bug,
the ``Platform'' used, and the ``Crash Type''
of the bug. To aid bug reproduction, the report indicates which ``Sanitizer''
was used, the range and time window of commits where the software ``Regressed,'' and a
``Reproducer Testcase'' to trigger the bug. After ClusterFuzz verifies
that the bug is fixed, it posts a comment indicating the commit range
where the software was fixed. Sheriffbot tracks disclosure deadlines
and posts comments to warn about approaching deadlines (if a bug is still
unfixed) or notify that a bug passed a disclosure deadline.

We use Selenium~\cite{selenium}, a browser automation tool,
with Google Chrome to scrape OSS-Fuzz bug reports on Monorail.
We ethically scrape data in accordance with Monorail's
robots.txt file~\cite{monorail-robots-txt}.
We extract data fields from the bug reports' text via pattern matching.
We scrape 23,907 bug reports from 316 projects, with report dates spanning
from May 2016 to October 2020\footnote{due to OSS-Fuzz's disclosure policy,
not all bug reports from July--October 2020 were publicly available at the time of data collection.}.

\section{Fault Characteristics}
\label{sec:fault-characteristics}

To gain a better sense of the landscape of OSS-Fuzz bugs, we
begin by examining the following fault characteristics:

\begin{description}
	\item [Fault type] A categorization of faults; e.g., timeout, out of memory, null dereference.
	\item [Flakiness] Whether a bug is reliably reproducible.
	\item [Fuzz blocker] Whether a fuzzer encounters the same bug very often,
	which blocks further fuzzing of downstream code.
	\item [Unfixedness] Whether a bug is unfixed.
	\item [CVE] Whether a bug has an associated record in the Common Vulnerabilities
	and Exposures (CVE) system~\cite{cve}.
\end{description}

The remainder of the section motivates, describes, and analyzes these characteristics
individually, and examines relationships between them.

\begin{figure}
	\centering
	\includegraphics[width=\columnwidth]{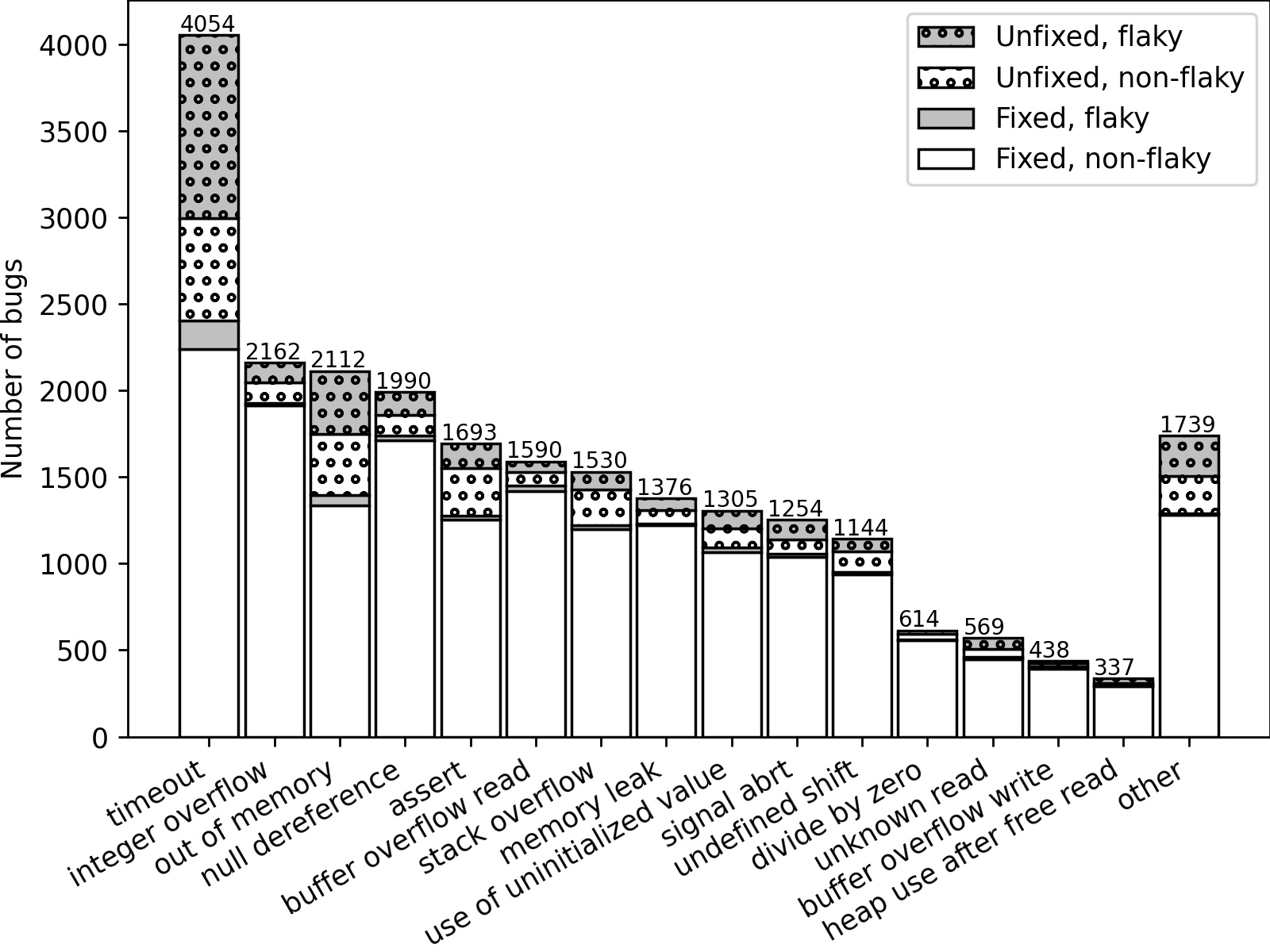}
	\caption{Numbers of bugs among the top 15 fault types.}
	\label{fig:fault-types}
\end{figure}

\subsection{Fault types}
\label{sec:types}

Fuzzing is often discussed in the context of software security as
an effective tool for uncovering security vulnerabilities~\cite{sp05-violate-assumptions,
takanen-fuzzing-book, mayhem, oss-fuzz, google-why-fuzz, cacm20-fuzzing,
icmins12-vuln-discovery-survey, li-fuzzing-survey, zeller-fuzzing-book,
blackhat05-file-fuzz},
particularly in finding buffer and numeric overflows --- two of the most
common software security vulnerabilities~\cite{li-sec-patches,
liu-vuln-distribution, shahzad-vuln-lc, cve-vuln-types}.
Prior advances in fuzzing research targeted specific fault types, such
as timeouts~\cite{slowfuzz, perffuzz, hotfuzz},
out of memory errors~\cite{fuzzfactory, memlock},
integer overflows~\cite{asplos15-int-overflows},
or buffer overflows~\cite{dowser, ietsoft16-mouzarani, secdev20-qasan, usenixsec20-jia}.
The attention on fuzzing as a security testing technique and
prior work on targeting specific fault types motivates the following question:

\begin{description}
	\item [RQ-FT] Which fault types do OSS-Fuzz's
	coverage-guided fuzzers frequently find?
\end{description}

\paragraph{Methodology}
To determine fault type, we use the ``Crash Type''
field in bug reports (e.g., in Figure~\ref{fig:bug-report}, the
``Crash Type'' is ``Unsigned-integer-overflow'').
We standardize some text (e.g., group together ``timeout'' and ``hang'',
or ``null dereference'' and ``null reference''), and
we consolidate heap, stack, and global overflows and underflows
into buffer overflow.
We group together ``null dereference read'' and
``null dereference write,'' since reading or writing to a null address
usually have similar consequences. For the opposite reasons,
we distinguish between ``buffer overflow read'' and ``buffer overflow write,''
since overreads primarily threaten confidentiality, while overwrites
also threaten integrity and can lead to arbitrary code execution.

\paragraph{Results}
Figure~\ref{fig:fault-types} shows the number of bugs among the top 15
fault types. Six of the most common fault types comprising
52\% \parentheticalstats{12316/23907} of bugs --- timeout, out of memory, null dereference,
stack overflow, memory leak, and signal abrt ---
primarily harm availability by crashing.
While such crashes might facilitate other attacks that compromise
confidentiality or integrity by, for example, exposing potential vulnerabilities
in the error-handling process, such crash-inducing faults are likely
less severe in their own capacity.

\takeaway{The majority of fuzzer-found bugs primarily harm availability.}

Four fault types comprising 23\% \parentheticalstats{5613/23907} of bugs ---
integer overflow, assertion violation, undefined shift, and divide by zero
--- indicate unintended program logic.
Such bugs can be exploited, for example, if an integer overflow affects
a buffer index and thus can facilitate a buffer overflow.
However, such unintended logic can also result in no more than abnormal
termination or bad output.

Three fault types comprising 14\% \parentheticalstats{3232/23907} of bugs
--- buffer overflow read, use of uninitialized value, and heap use after free read ---
do primarily jeopardize memory confidentiality.
Buffer overflow writes \parentheticalstats{2\%, 438/23907} jeopardize memory integrity
and can lead to arbitrary code execution.

\subsection{Flakiness}

Fuzzers can sometimes find flaky bugs, where a test input cannot
reliably reproduce a bug. ClusterFuzz --- the continuous fuzzing
engine behind OSS-Fuzz --- usually ignores unreproducible
bugs; however, if a flaky bug appears very frequently, then
ClusterFuzz will file a bug report~\cite{clusterfuzz-fixing-a-bug}.
Such flakiness motivates the following research questions:

\begin{description}
	\item [RQ-FLK] How prevalent are flaky bugs?
	\item [RQ-FLK-FT] Which fault types are disproportionately flaky?
\end{description}

\paragraph{Methodology}

To identify flaky bugs, we look for bugs which ClusterFuzz
deems insufficiently reproducible, or where there appears a comment in
the bug report including phrases suggesting irreproducibility
(e.g., ``unreproducible,'' ``can't reproduce'').
This produces a conservative count of flaky bugs, since ClusterFuzz only
reports a flaky bug if it appears very often, and developers
experiencing reproducibility problems may stay silent, complain
outside of the bug report's comment section on Monorail (e.g., they may
use their project's own issue tracker), or use different phrasing to suggest flakiness.

\paragraph{Results}
Out of 23907 studied bugs, we identify 3139 \parentheticalstats{13\%} as flaky.
Figure~\ref{fig:fault-types} breaks down the composition of flaky and non-flaky
among the top 15 fault types. We observe high rates of
flakiness in timeouts \parentheticalstats{30\%, 1225/4054, $p < 10^{-272}$ via a $\chi^2$ test on
the null hypothesis that timeouts and non-timeouts are equally flaky}
and out of memory errors \parentheticalstats{20\%, 424/2112, $p < 10^{-22}$ via a $\chi^2$ test on the null hypothesis
that OOMs and non-OOMs are equally flaky}. Since both timeouts and OOMs
are resource exhaustion bugs, the unpredictability of resource availability and usage
may hamper reproducibility.

\takeaway{Timeouts and OOMs are
disproportionately flaky.}

\begin{figure}
	\centering
	\includegraphics[width=\columnwidth]{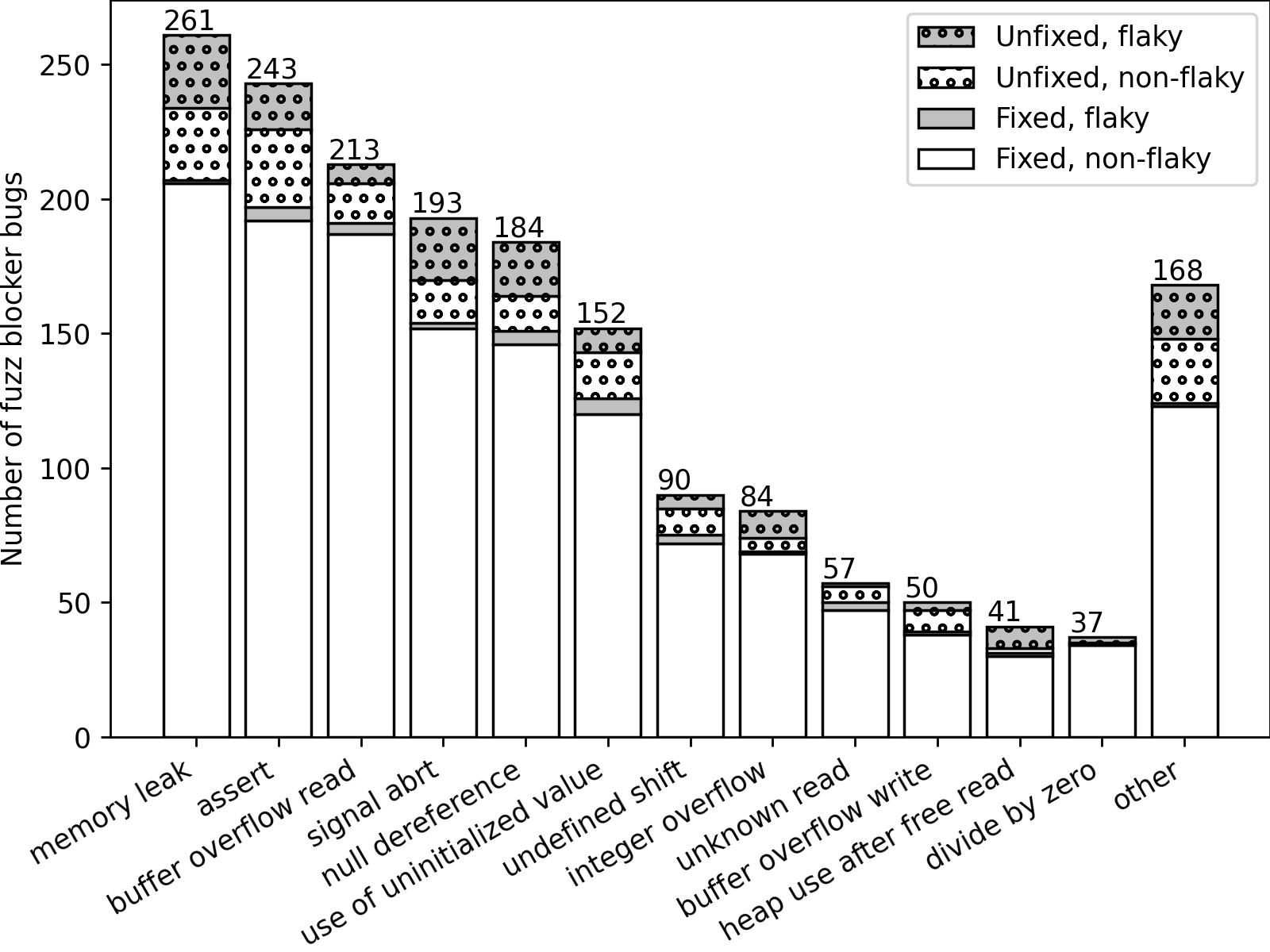}
	\caption{Numbers of fuzz blockers among the top 12 fault types.}
	\label{fig:fuzz_blocker_fault_types}
\end{figure}

\subsection{Fuzz blockers}
\label{sec:fuzz-blockers}

If a bug occurs very frequently, it
can block subsequent fuzzing, hurting fuzzer performance.
ClusterFuzz reports these frequently
crashing bugs as \emph{fuzz blockers}, and advises maintainers that fixing
such bugs would lead to better fuzzing.
The fuzzing community expressed concern that while fuzz blockers are
important to fix from a fuzzing practitioner's viewpoint, developers may
ascribe low priority to fuzz blockers~\cite{fuzzcon20-holler}.
We thus ask the following research questions:

\begin{description}
	\item [RQ-BLK] How prevalent are fuzz blockers?
	\item [RQ-BLK-FC] What relationships exist between fuzz blocker prevalence
	and other fault characteristics?
\end{description}

\paragraph{Methodology}
If ClusterFuzz detects that a bug appears very frequently, then ClusterFuzz
adds a ``Fuzz-Blocker'' label to the bug report.
We use ClusterFuzz's labeling.
However, based on ClusterFuzz's source code~\cite{clusterfuzz-src-appengine-handlers-cron-cleanup-py},
we notice that ClusterFuzz does not report fuzz blocking timeouts, out of memory errors,
or stack overflows. We speculate that ClusterFuzz's developers excluded these fault types
since they block fuzzing so often that distinguishing blockers from non-blockers is uninteresting.
Given the lack of data on the excluded fault types, we do not count them as fuzz blockers.

\paragraph{Results}
Out of 16211 bugs that ClusterFuzz examined for fuzz blockage,
1773 \parentheticalstats{11\%} are fuzz blockers, of which
10\% \parentheticalstats{185/1773} are flaky.
Figure~\ref{fig:fuzz_blocker_fault_types} shows the 12 most prevalent
fault types among ClusterFuzz-identified fuzz blockers.
Memory leaks are the most common fault type among fuzz blockers
\parentheticalstats{15\%, 261 of 1773 fuzz blockers}, and fuzz blocking
memory leaks comprise 19\% \parentheticalstats{261/1376} of all memory leaks.
Leak detection is part of ASan~\cite{asan}, but can be disabled
to ignore memory leaks and continue fuzzing code downstream of any leaks.
Assertion violations are the second most common fault type among fuzz blockers
\parentheticalstats{14\%, 243 of 1773 fuzz blockers}, and fuzz blocking
assertion violations comprise 14\% \parentheticalstats{243/1693} of all
assertion violations. Recompiling code without the fuzz blocking assertion would
similarly unblock fuzzing downstream of the assertion failure.

\takeaway{Memory leaks and assertion violations are the most common ClusterFuzz-identified fuzz blockers,
both of which can be disabled.}

Although the top two fault types are relatively low-impact, buffer overflow read,
a high severity fault type, is the third most common fault type among ClusterFuzz-identified fuzz blockers
\parentheticalstats{12\%, 213 of 1773 fuzz blockers}, and comprises
13\% \parentheticalstats{213/1590} of all buffer overflow reads.
Since fuzz blockers appear very often during fuzzing, an attacker may find
a fuzz blocking buffer overflow quickly (we confirm the intuition that fuzz
blockers are quickly found in Sec.~\ref{sec:time-to-detect}:~\nameref{sec:time-to-detect}).

\subsection{Unfixed bugs}

Developers may choose not to fix a bug for many reasons. A bug might
be too hard to repair (e.g., if a bug is hard to replicate, or requires an
expensive code overhaul). Maintainers might also judge that a reported bug
is not actually a bug or is not important enough to warrant a fix
(e.g., if developers think a fuzzer-found error case is low-impact, obscure, or
will not manifest in practice). Fuzzers, in particular, can flood software
projects with many seemingly low-impact bugs~\cite{z3-fuzz-argument, fuzzcon20-holler}.
Attackers, however, can use known bugs as a low-cost starting point to craft
attacks with existing vulnerabilities, gather intelligence to eventually craft a
more complex exploit, or gauge the agility of a software team's response to faults.
We thus ask:

\begin{description}
	\item [RQ-NOFIX] How many fuzzing-identified bugs are unfixed?
	\item [RQ-NOFIX-FC] Which bugs are often unfixed?
\end{description}

\paragraph{Methodology and Results}
We examine publicly visible bugs without a verified fix at the time of data collection.
Of the publicly visible bugs, 22\% \parentheticalstats{5148/23907} are unfixed.
OSS-Fuzz does not normally publicize unfixed bugs until after the 90-day disclosure deadline,
except for 130 unfixed bugs that were publicized early by developers and whose data we collected
prior to the default 90-day deadline. Bugs unfixed at the time of data collection were left
unfixed for a median of 437 days, with 95\% of bugs unfixed for 90--1275 days.

Figure~\ref{fig:fault-types} breaks down the composition of fixed and
unfixed bugs according to flakiness and fault types.
Flaky bugs, even if appearing frequently, are overwhelmingly
not fixed \parentheticalstats{86\%, 2684/3139}. Non-flaky bugs are unfixed only
12\% \parentheticalstats{2464/20768} of the time.
We postulate that some flaky bugs, even though they appear frequently, are not actually faults in the software itself;
for example, software might timeout due to a temporarily unavailable resource.
Irreplicability also hinders attempts to diagnose and understand the bug.
Although OSS-Fuzz encourages developers to write speculative patches
for irreplicable bugs based on the information in the bug report,
developers may hesitate to write speculative patches with limited information
for a fault that may not actually exist.

\takeaway{Flaky bugs are overwhelmingly unfixed.}

Because flaky bugs are often unfixed, we exclude them when
comparing fix rates among different fault types to prevent flakiness
from acting as a confounding variable.
Three fault types have disproportionately many unfixed bugs compared to other fault types:
timeout \parentheticalstats{21\%, 589 of 2829 non-flaky bugs are unfixed, $p<10^{-55}$
with a $\chi^2$ test on the null hypothesis that timeouts and non-timeouts
have the same frequency of unfixed bugs},
out of memory \parentheticalstats{21\%, 353/1688, $p<10^{-32}$}, and
assertion violation \parentheticalstats{18\%, 274/1530, $p<10^{-13}$}.

However, as discussed in Section~\ref{sec:types}, 
timeouts and out of memory errors primarily
hamper availability rather than confidentiality or integrity, and
assertion violations do not necessarily pose an immediate threat to
security or reliability. Considering the generally lower security impact of such
bugs, the lower fix rate suggests greater developer apathy towards
such bugs. Timeouts and out of memory errors may
also be more annoying for developers to reproduce and fix, as reproducing
such bugs consume substantial time and/or hardware resources, which
slows the potentially repetitive process of analyzing the bug and
testing patches. The lower impact and greater pain in fixing such bugs
may explain the lower fix rate.

\takeaway{Timeouts, out of memory errors,
and assertion violations are more frequently unfixed compared to other fault types, even if not flaky.}

\subsection{CVEs}
\label{sec:cves}

The Common Vulnerabilities and Exposures (CVE) list~\cite{cve}
is a public reference for security vulnerabilities.
Organizations designated as CVE Numbering Authorities (e.g., MITRE, Debian,
Microsoft, PHP Group) can issue CVEs for vulnerabilities discovered
in-house or reported from third parties.
Various security tools, such as threat intelligence dashboards
and security scanning tools, use CVEs as a source of threat information.
The security community expressed concern that
very few of OSS-Fuzz's security vulnerabilities were issued CVEs~\cite{oss-fuzz-no-cves},
and thus most of OSS-Fuzz's discovered vulnerabilities are
not visible to security tools that rely on CVEs.
This prompts the following research questions:

\begin{description}
	\item [RQ-CVE] How many OSS-Fuzz bugs have CVE records?
	\item [RQ-CVE-FT] Which fault types often receive CVEs?
\end{description}

\paragraph{Methodology and Results}
A CVE record usually references bug reports or other
documentation on a vulnerability. We mine CVE records~\cite{cve-csv} to find URL
references to OSS-Fuzz bug reports. We produce a conservative list of OSS-Fuzz
bugs with CVEs, since a CVE's list of references may be incomplete and omit a
reference to an OSS-Fuzz bug report.

We find 98 OSS-Fuzz issues with CVEs, a small number
relative to the over 20,000 bugs found by OSS-Fuzz.
Table~\ref{tab:cves} presents the fault types of bugs with CVEs.
Most bugs with CVEs are dangerous memory operations
that threaten confidentiality or integrity, although people have also filed CVEs for
bugs that primarily affect availability, such as timeouts, out of memory
errors, and null dereferences.

\takeaway{Few OSS-Fuzz bugs result in CVEs, and most filed CVEs are for
memory corruption bugs.}

\begin{table}
\centering
\begin{tabular} {l r}
	\toprule
	Fault type & OSS-Fuzz bugs with filed CVEs \\
	\midrule
	dangerous memory write\tablefootnote{Buffer overflow write (40), heap use after free write (2), unknown write (2),
	stack use after return write (1), container overflow write (1).} & 46 \\
	dangerous memory read\tablefootnote{Buffer overflow read (16), unknown read (8), heap use after free read (6),
	use of uninitialized value (4).} & 34 \\
	null dereference & 4 \\
	out of memory & 4 \\
	heap double free & 2 \\
	timeout & 2 \\
	other\tablefootnote{Divide by zero (1), signal abrt (1), negative size param (1), floating point exception (1),
	assert (1), memory leak (1).} & 6 \\
	\bottomrule
\end{tabular}
	\caption{Number of CVEs filed per fault type. Most CVEs are filed for memory corruption bugs.}
	\label{tab:cves}
\end{table}

\paragraph{Limitations}
Outsiders might independently co-discover and file CVEs for vulnerabilities
found by OSS-Fuzz. Thus, the actual number of CVEs filed by project members
in response to an OSS-Fuzz discovery is likely even lower.

\section{Fault Lifecycle}
\label{sec:fault-lifecycle}

Prior work~\cite{li-sec-patches} found that the median
lifespan --- the time between fault introduction and repair ---
of vulnerabilities was 438 days.
Numeric errors and buffer overflows specifically had median lifespans
of 659.5 and 781 days respectively. In that time, an attacker may
find and exploit such vulnerabilities. Continuous fuzzing aims to
shorten bug lifespans and help developers stay ahead of attackers.
We study the following aspects of OSS-Fuzz bugs' fault lifecycles:

\begin{description}
	\item [Time-to-detect] The time from fault introduction to detection.
	\item [Time-to-fix] The time from fault detection to repair.
\end{description}

The remainder of the section examines these aspects and their relationships
to fault characteristics discussed in Sec.~\ref{sec:fault-characteristics}.

\subsection{Time-to-detect}
\label{sec:time-to-detect}

By continuously probing software for faults, continuous fuzzing aims to
shorten the time to detect regressions.
Heartbleed, a buffer overflow read vulnerability in
OpenSSL, was unfixed for two years~\cite{heartbleed},
but is now a demonstrative bug for fuzzing~\cite{libfuzzer-tutorial}.
Continuous testing techniques, such as continuous fuzzing, can shorten
the lifespan of bugs and security vulnerabilities through rapid
fault detection.
Such lengthy bug lifespans, and the prospect of shortening such lifespans
via continuous fuzzing, motivate the following research question:

\begin{description}
	\item [RQ-T2D] How much time elapses between fault introduction
	and detection (the \emph{time-to-detect})?
\end{description}

We also probe the relationship between fault characteristics and time-to-detect
to reveal, for example, whether bugs of certain fault types take longer to detect.

\begin{description}
	\item [RQ-T2D-FC] What relationships exist between fault characteristics
	and time-to-detect?
\end{description}

\paragraph{Methodology}
ClusterFuzz, after finding a bug, identifies
the range of commits where the regression was introduced.
We use the right time boundary of the regression range as the time of fault
introduction. In the example presented in Figure~\ref{fig:bug-report},
the time of fault introduction (2019-12-19-03:18 UTC) is the second datetime (201912190318)
of the ``range'' value of URL in the ``Regressed'' field.
We use the time of bug reporting as the time of bug detection;
in Figure~\ref{fig:bug-report}, the time of bug reporting is ``Fri, Jan 10, 2020 7:15 AM EST.''
The time-to-detect is the time elapsed between these two times.

\begin{figure}
	\centering
	\includegraphics[width=\columnwidth]{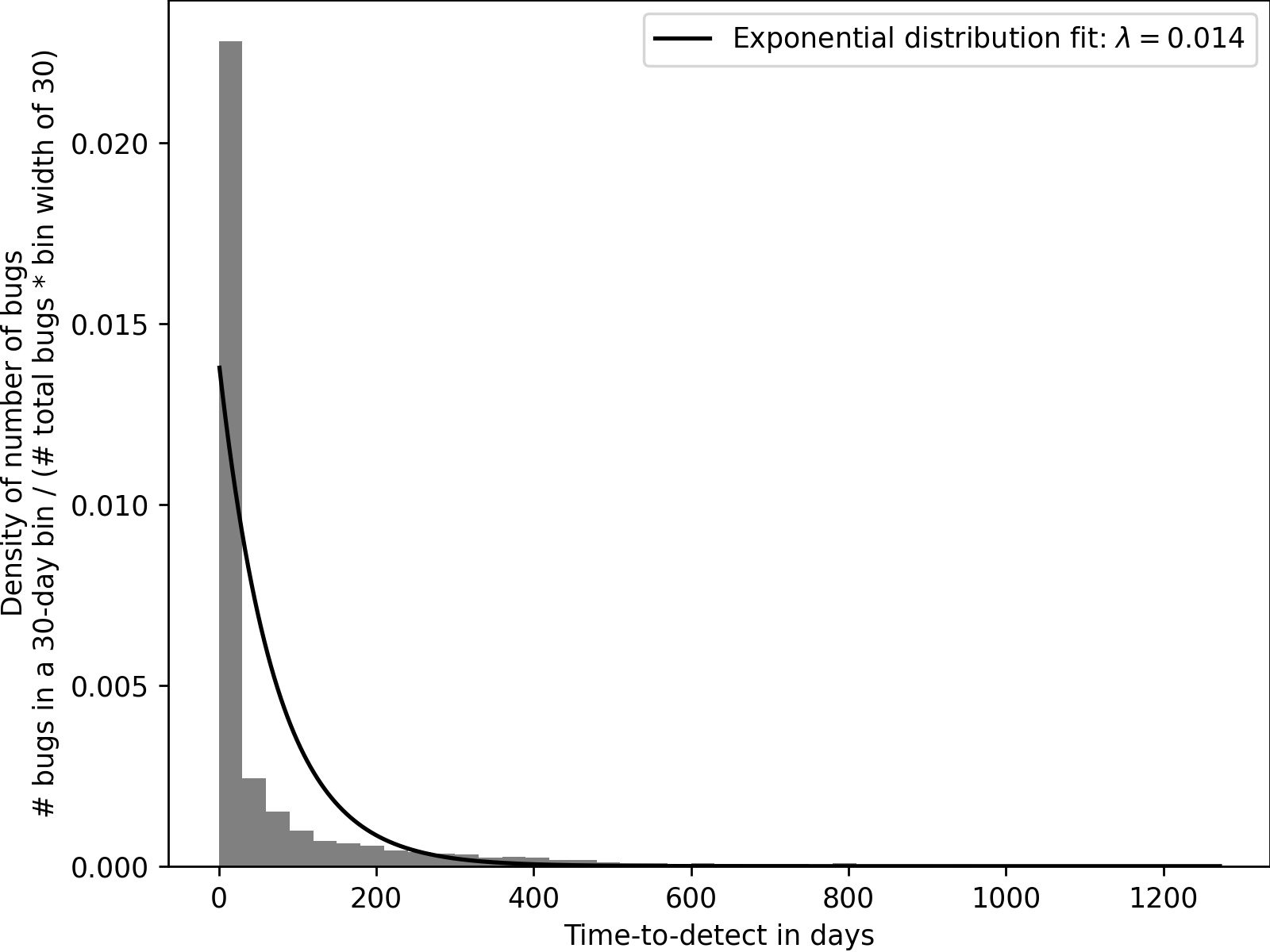}
	\caption{Density of bugs with respect to time-to-detect.
		Bugs are often detected quickly, with the time intervals following
		an exponential distribution.}
	\label{fig:time_to_detect_distribution}
\end{figure}

\begin{figure}
	\begin{subfigure}[b]{\columnwidth}
		\centering
		\includegraphics[width=\columnwidth]{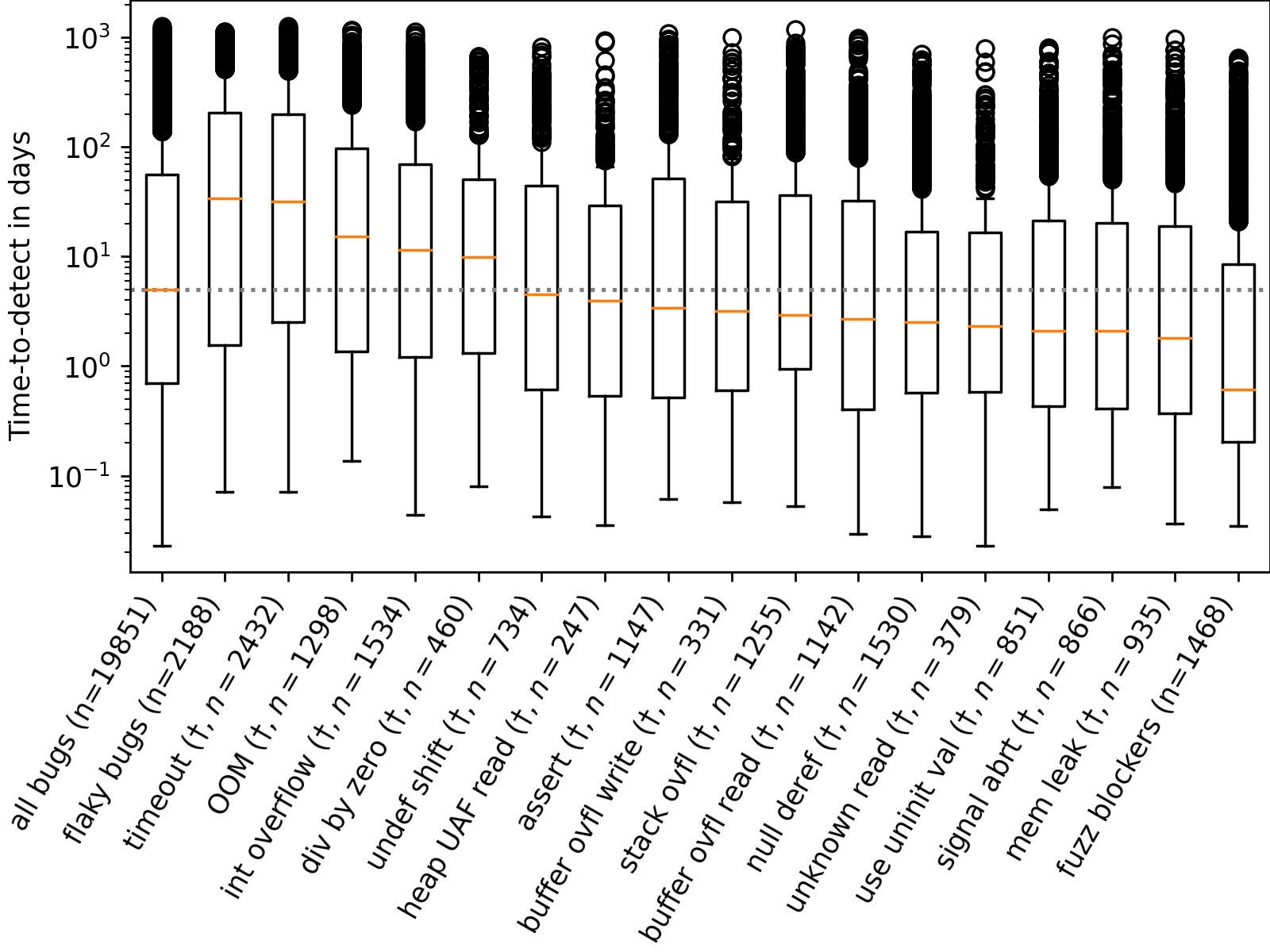}
		\caption{Times-to-detect among flaky bugs, fuzz blockers, and
		the 15 most common fault types.
		The dotted line is the median time over all analyzed bugs.
		Flaky bugs and multiple fault types take significantly longer to detect.
		Fuzz blockers and multiple fault types take significantly shorter to detect.}
		\label{fig:time_to_detect_boxplots}
	\end{subfigure}
	\begin{subfigure}[b]{\columnwidth}
		\centering
		\includegraphics[width=\columnwidth]{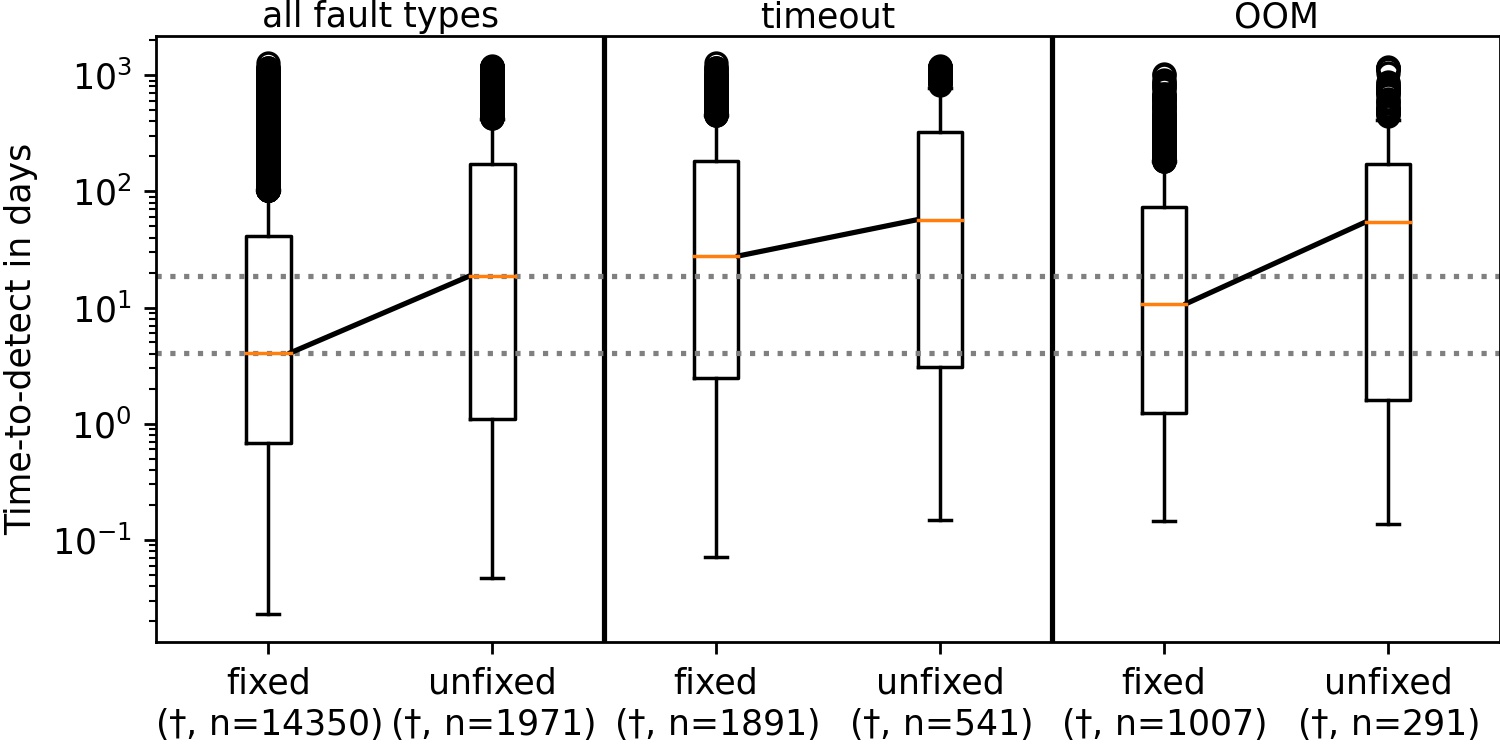}
		\caption{Comparison of times-to-detect of bugs that developers (do not) fix.
		The lower (resp. upper) dotted line is the median time-to-detect over all
		fixed (unfixed) non-flaky/blocker bugs.
		Unfixed bugs generally have longer times-to-detect.}
		\label{fig:time_to_detect_boxplots_isfixed}
	\end{subfigure}
	\caption{Times-to-detect among bugs with various characteristics.
	A $\dagger$ means flaky bugs and fuzz blockers are excluded.}
\end{figure}

\paragraph{Results}
Figure~\ref{fig:time_to_detect_distribution} shows the density of bugs
over the time-to-detect axis. The density of bugs in a histogram bin is the
number of bugs in the bin normalized by the width of the bin and the
total number of bugs shown in the histogram. Such normalization makes the
areas of the bars sum to one, akin to how the area under
a probability density function sum to one (100\% probability).
The time-to-detect a bug is exponentially
distributed \parentheticalstats{$p < 10^{-308}$ via a Kolmogorov-Smirnov goodness of fit test}.
OSS-Fuzz's fuzzers find most regressions quickly, with a median time-to-detect of 5 days.
Our findings complement, with in-the-wild data, prior findings on the exponential
cost of fuzzing~\cite{esecfse20-exp-cost}. Since coverage-guided fuzzing is a
guided search through a combinatorially large search space of inputs, and fuzzer coverage
plateaus over time~\cite{zeller-fuzzing-book}, the exponential
distribution is expected.
The short time-to-detect provides further evidence that coverage-guided fuzzing is
well-suited for continuous testing.

\takeaway{OSS-Fuzz detects the majority of identified regressions within a week.}

Figure~\ref{fig:time_to_detect_boxplots} compares the time-to-detect among
various characteristics. Flaky bugs take much longer to detect than
non-flaky bugs, with medians of 34 vs. 4 days to detect
flaky vs. non-flaky bugs \parentheticalstats{$p < 10^{-109}$ via a two-sided Mann-Whitney U test
on the null hypothesis that flaky and non-flaky bugs have the same time-to-detect}.
On the other hand, ClusterFuzz-identified fuzz blockers
are found much earlier than non-blockers \parentheticalstats{medians of 0.6 vs. 3.3 days, $p < 10^{-82}$,
excluding the fault types that ClusterFuzz do not report fuzz blockers on}.
Since fuzz blockers appear very frequently by definition,
such bugs would appear more often in the search space, which increases the
probability of discovering a fuzz blocker early.

\takeaway{OSS-Fuzz finds flaky bugs late, and ClusterFuzz-identified fuzz blockers early.}

The respectively long and short times-to-detect of flaky bugs and fuzz blockers prompt us to
exclude these bugs when comparing fault types to prevent flakiness or fuzz blockers
from acting as confounding variables.
For example, flaky bugs, if not excluded, would skew timeout time-to-detect upwards due
to the high prevalence of flaky timeouts.
The median time-to-detect non-flaky, non-blocker bugs is 4.8 days.

Timeouts \parentheticalstats{median time-to-detect of 32 days, $p < 10^{-158}$ via two-sided U test on
the null hypothesis that timeouts and non-timeouts have the same time-to-detect}
and out of memory errors \parentheticalstats{15 days, $p < 10^{-25}$} take
significantly longer than other bug types to detect.
One possibility is that timeouts and OOMs can result from runaway looping or recursion.
Fuzzers' coverage metrics often do not account for loop iterations or recursion depth,
which reduces the benefit of coverage in guiding the input search towards inputs that
consume a lot of resources.
Specialized fuzzers such as SlowFuzz~\cite{slowfuzz}, which optimizes for path length,
PerfFuzz~\cite{perffuzz}, which optimizes for basic block execution counts, or
\texttt{mem} (in FuzzFactory~\cite{fuzzfactory}), which optimizes for memory allocations,
can guide the search for such bugs more effectively and reduce the time-to-detect.

\takeaway{Timeouts and out of memory errors take longer to detect.}

Integer overflows \parentheticalstats{12 days, $p < 10^{-13}$}
and divide by zeroes \parentheticalstats{10 days, $p = 0.0007$} also take longer to detect.
Meanwhile, memory leaks \parentheticalstats{1.8 days, $p < 10^{-20}$},
signal abrt \parentheticalstats{2.1 days, $p < 10^{-14}$},
use of uninitialized values \parentheticalstats{2.1 days, $p < 10^{-13}$},
unknown reads \parentheticalstats{2.3 days, $p < 10^{-5}$},
and null dereferences \parentheticalstats{2.5 days, $p < 10^{-22}$} are faster to detect.
Some of the variance between fault types may be the result of ClusterFuzz's
prioritization of ASan, which detects memory access violations and leaks, and MSan,
which detect use of uninitialized values, over UBSan, which detects undefined behavior
such as integer overflow or divide by zero~\cite{blackhat19-clusterfuzz}.

The three longest to detect fault types --- timeout, OOM, and integer overflow
--- are by no means sparse; they are the three most prevalent fault types (Section~\ref{sec:types}).
Despite the extra time taken to find such bugs, they are plentifully discoverable.


\takeaway{Slower-to-detect fault types are not sparse.}

Figure~\ref{fig:time_to_detect_boxplots_isfixed} compares times-to-detect among
fixed and unfixed bugs. We again exclude flaky bugs or fuzz blockers to avoid
confoundment, especially since flaky bugs are largely unfixed. Unfixed bugs often
take longer for fuzzers to detect \parentheticalstats{$p < 10^{-55}$ via a two-sided U test
on the null hypothesis that fixed and unfixed bugs have equal times-to-detect}.
Perhaps the longer times-to-detect of unfixed bugs is indicative of fault complexity
that makes the bugs both harder for fuzzers to find (hence the long time-to-detect)
and harder for people to fix (hence the bug is unfixed).
Perhaps developers are also more likely to forget about the details of an older code change
and neglect to fix an older bug.


\takeaway{Unfixed bugs take longer to detect.}

Since timeouts and out of memory errors are disproportionately unfixed, and
unfixed bugs take longer to detect, we examine (un)fixed bugs among these
two fault types to gauge the influence of unfixed bugs on the long times-to-detect
of these two fault types.
While unfixed bugs do take longer to detect among bugs of both fault types
\parentheticalstats{$p < 10^{-5}$ via a two-sided U test for timeouts, $p < 10^{-6}$ for OOMs},
fixed timeouts take longer to detect than fixed non-timeouts \parentheticalstats{$p < 10^{-125}$},
and unfixed timeouts take longer to detect than unfixed non-timeouts \parentheticalstats{$p < 10^{-16}$}.
The same pattern appears in OOMs \parentheticalstats{$p < 10^{-17}$ comparing fixed OOMs vs. non-OOMs,
$p = 0.003$ comparing unfixed OOMs vs. non-OOMs}.
Thus, timeouts and OOMs' long times-to-detect are likely also attributable to factors
other than those driving an increase in time-to-detect among unfixed bugs.

\subsection{Time-to-fix}
\label{sec:repair-urgency}

Continuous fuzzing is most effective if developers promptly fix bugs.
Otherwise, an accumulation of bugs can leave openings for attackers and
hamper discovery of more bugs downstream of the unfixed faults.
To promote prompt repair, OSS-Fuzz applies a 90-day disclosure policy.
The severity, ease of repair, and developer attitudes toward bugs may affect the
time to fix bugs. Prior work found that developers fix severe bugs almost twice
as fast~\cite{wcre12-time2fix}. Moreover, fuzzing campaigns can generate an
overload of low-priority bug reports~\cite{z3-fuzz-argument,fuzzcon20-holler}, suggesting a
need to direct fuzzing efforts towards high-value bugs that developers eagerly fix.
We thus ask:

\begin{description}
	\item [RQ-T2F] How much time elapses between fault detection and repair
	(the \emph{time-to-fix})?
	\item [RQ-T2F-FC] What relationships exist between fault characteristics
	and time-to-fix?
\end{description}

\paragraph{Methodology and Results}

\begin{figure}
	\centering
	\includegraphics[width=\columnwidth]{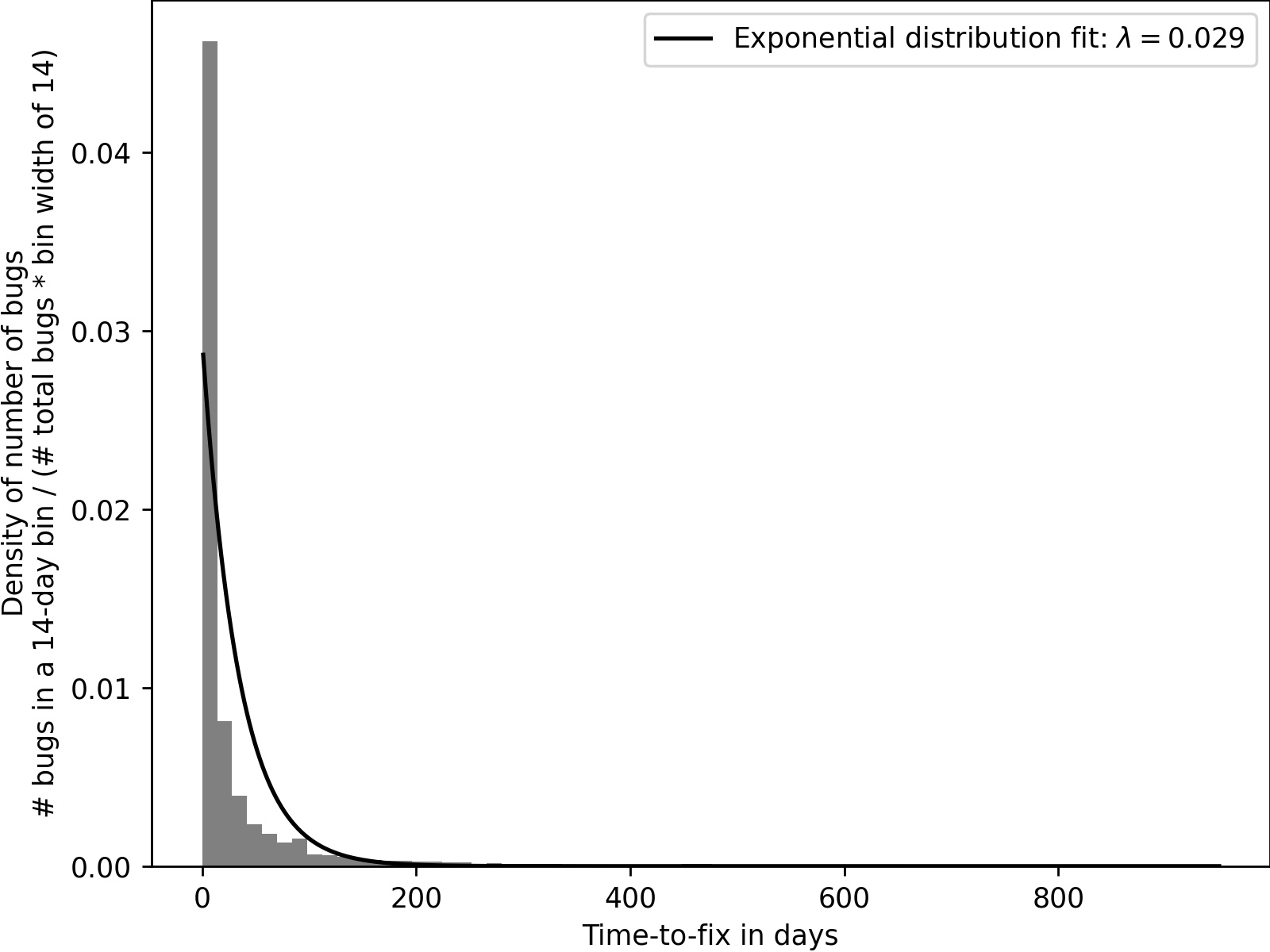}
	\caption{Density of bugs with respect to time-to-fix.
		Bugs are often fixed quickly, with the time intervals following
		an exponential distribution.}
	\label{fig:time_to_fix_distribution}
\end{figure}

\begin{figure}
	\centering
	\includegraphics[width=\columnwidth]{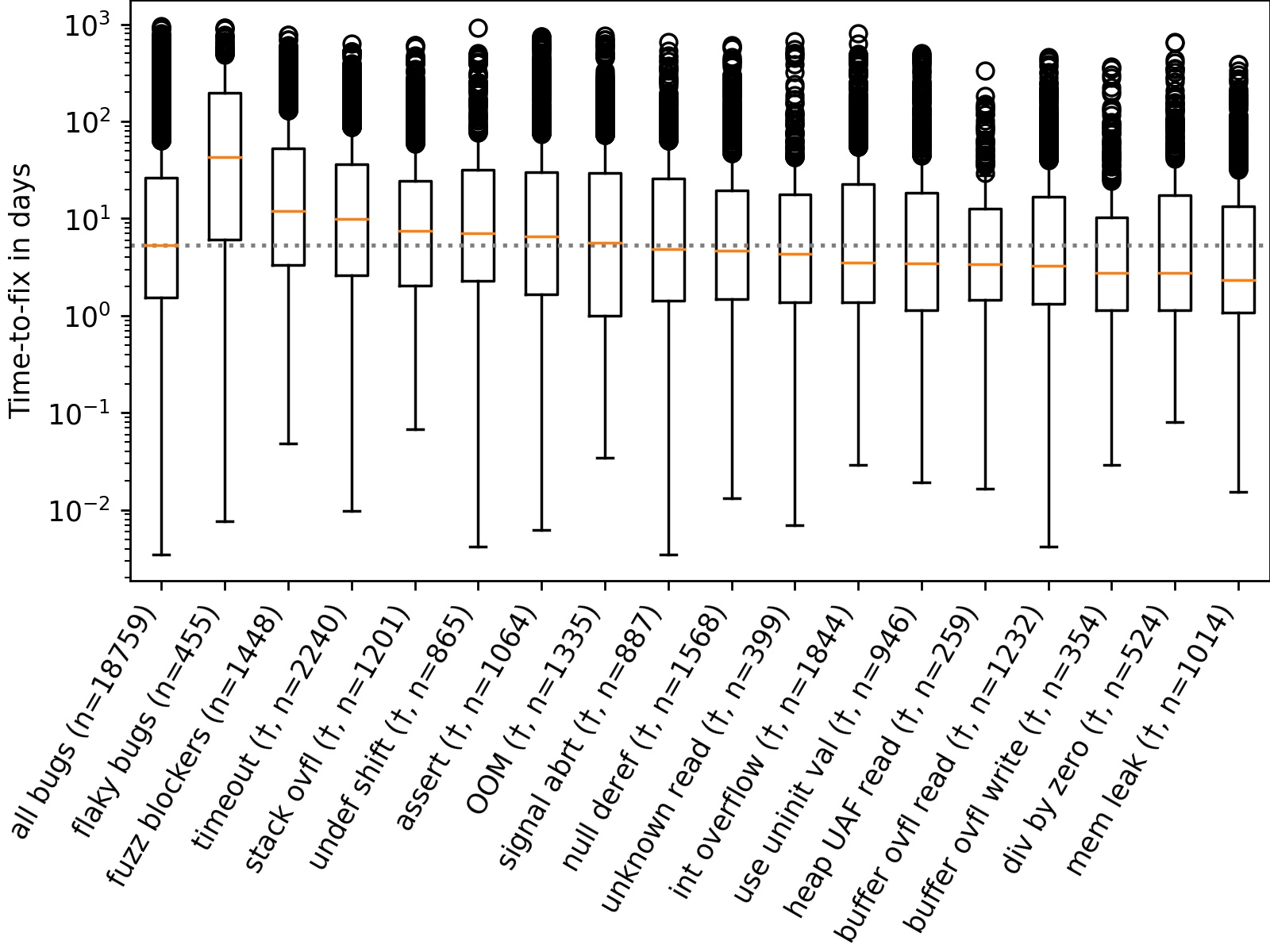}
	\caption{Times-to-fix.
	A $\dagger$ means flaky bugs and fuzz blockers were excluded.
	The dotted line is the median time over all analyzed bugs.}
	\label{fig:time_to_fix}
\end{figure}

We compute time-to-fix as the time from bug reporting to patch
verification (Comment 1 in Figure~\ref{fig:bug-report}).
Figure~\ref{fig:time_to_fix_distribution} shows the density of bugs
over the time-to-fix axis. The time-to-fix is exponentially distributed
\parentheticalstats{$p < 10^{-308}$ via a Kolmogorov-Smirnov goodness of fit test}.
Out of all fixed bugs, 90\% \parentheticalstats{16952/18759} are fixed within the 90-day
disclosure deadline; the median time-to-fix is 5.3 days.

\takeaway{Most bugs are repaired well within the 90-day disclosure period,
and over half are fixed within a week.}

Figure~\ref{fig:time_to_fix} compares the time-to-fix among
various bug categories.
Flaky bugs, the vast majority of which are already unfixed,
take an order of magnitude longer to repair, with a median time-to-fix
of 43 days, versus 5.1 days for non-flaky bugs \parentheticalstats{$p < 10^{-54}$ by a two-sided U test
on the null hypothesis that flaky and non-flaky bugs have the same time-to-fix}.
Given the difficulty of attempting to patch a hard-to-reproduce bug, the
very long time-to-repair is unsurprising and supports prior findings~\cite{msr14-works4me}
on bugs with reproducibility issues.

\takeaway{Flaky bugs, if fixed at all, are fixed very slowly.}

Since fuzz blockers impede fuzzing performance by blocking program exploration
downstream from the bug, quickly remedying blockers is important for a healthy
fuzzing campaign. We find, however, that developers fix ClusterFuzz-identified fuzz blockers
more slowly, with a median of 12 days as opposed to 4 for non-blockers
\parentheticalstats{$p < 10^{-63}$ via a two-sided U test,
excluding the fault types that ClusterFuzz do not report fuzz blockers on}.
Our finding confirms prior concerns~\cite{fuzzcon20-holler} on developers'
low prioritization of fuzz blockers, suggesting a need for greater awareness
on the need to address blockers.

\takeaway{Fuzz blockers are fixed less, rather than more, urgently.}

The long times-to-fix of flaky bugs and fuzz blockers prompt us to once again
exclude these bugs when comparing fault types to prevent flakiness or fuzz blockers
from acting as confounding variables.
Timeouts \parentheticalstats{median time-to-fix of 10 days, $p < 10^{-45}$ via two-sided U test on
the null hypothesis that timeouts and non-timeouts have the same time-to-fix},
stack overflows \parentheticalstats{7.5 days, $p < 10^{-5}$},
and undefined shifts \parentheticalstats{7 days, $p < 10^{-10}$}
have longer times-to-fix, while memory leaks \parentheticalstats{2.3 days, $p < 10^{-18}$}
and divide by zeroes \parentheticalstats{2.8 days, $p < 10^{-4}$} have
shorter times-to-fix.

Although these aforementioned fault types can indicate faulty logic
or impact availability via abnormal termination,
they do not entail obvious consequences for confidentiality or
integrity. Yet timeouts and memory leaks have the longest and shortest
times-to-fix respectively among the top 15 fault types. Severity
does not always correlate with the urgency of a patch.

\takeaway{Fault type severity does not always correlate with fix speed.}

The short times-to-fix of memory leaks and divide by zeroes may suggest
that such bugs are easier to fix, and thus fixed more quickly.
Meanwhile, timeouts and stack overflows, even if not flaky,
can be more annoying to repeatedly reproduce in the debugging process,
and may be more difficult to localize and repair, resulting in longer times-to-fix.

Buffer overflow writes \parentheticalstats{2.8 days, $p < 10^{-6}$} also have shorter times-to-fix.
Since such bugs are so severe --- an attacker might
execute arbitrary and malicious code --- we are encouraged to see urgency in
responding to such bugs.

\takeaway{Buffer overflow writes, which are very severe, are fixed more urgently.}


\paragraph{Limitations}
Some bugs have very short times-to-repair; 40 were fixed within one hour after
the bug was reported. While some of these bugs may be repaired very quickly due
to a rapid response to the bug report, we suspect that some bugs were discovered
and patched before OSS-Fuzz had produced a bug report. We are not aware of an
effective method to distinguish between the two cases, but the small number of
such bugs poses only a marginal threat to validity.

\section{Longitudinal Evolution}
\label{sec:longitudinal-evolution}

\newcommand\minifigcaption[1]{\captionsetup{aboveskip=0pt, belowskip=8pt}\caption{#1}}
\begin{figure*}
	\centering
	\begin{subfigure}[b]{.166\linewidth}
		\centering
		\includegraphics[width=\linewidth]{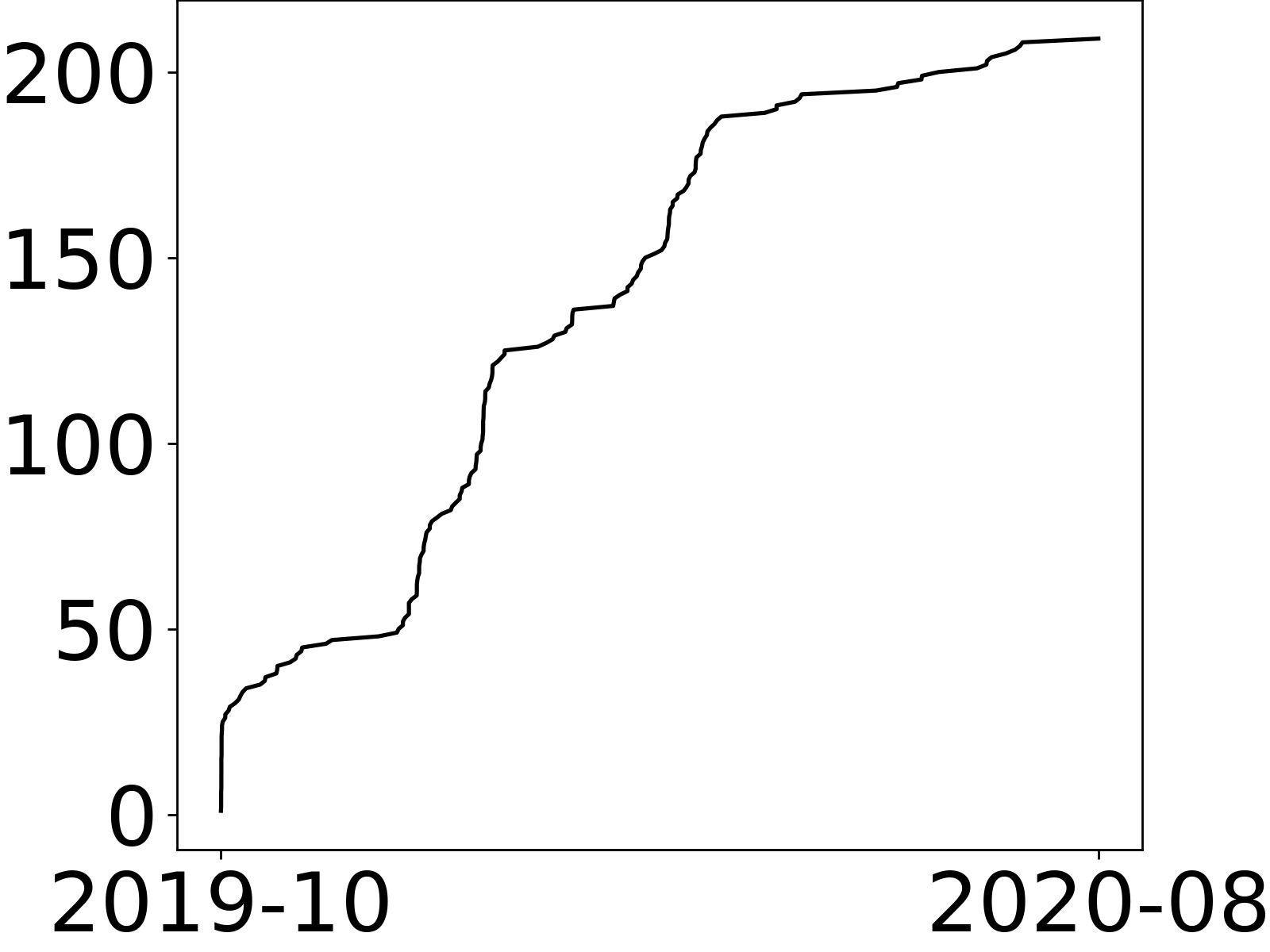}
		\minifigcaption{binutils}
	\end{subfigure}%
	\begin{subfigure}[b]{.166\linewidth}
		\centering
		\includegraphics[width=\linewidth]{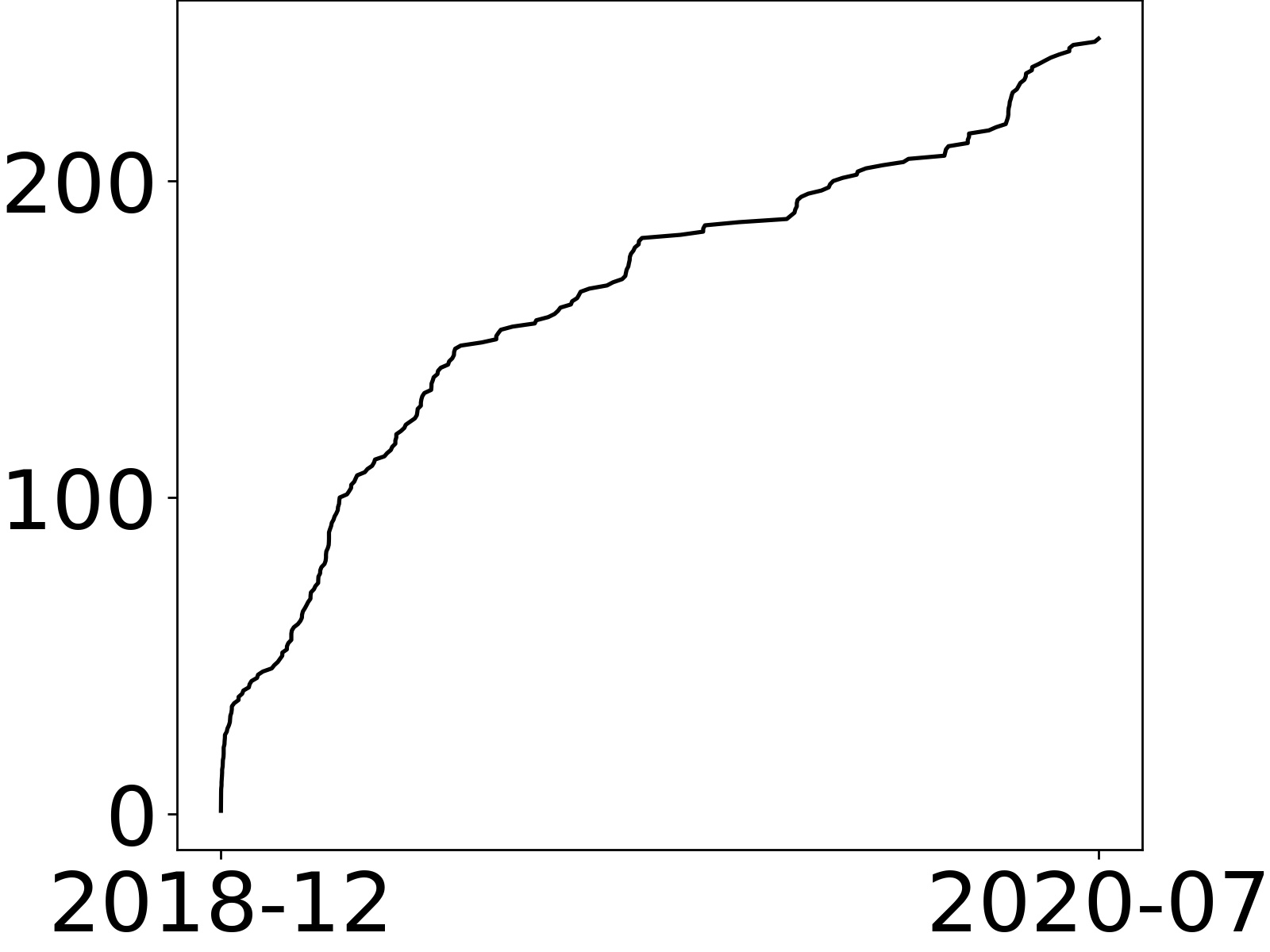}
		\minifigcaption{clamav}
	\end{subfigure}%
	\begin{subfigure}[b]{.166\linewidth}
		\centering
		\includegraphics[width=\linewidth]{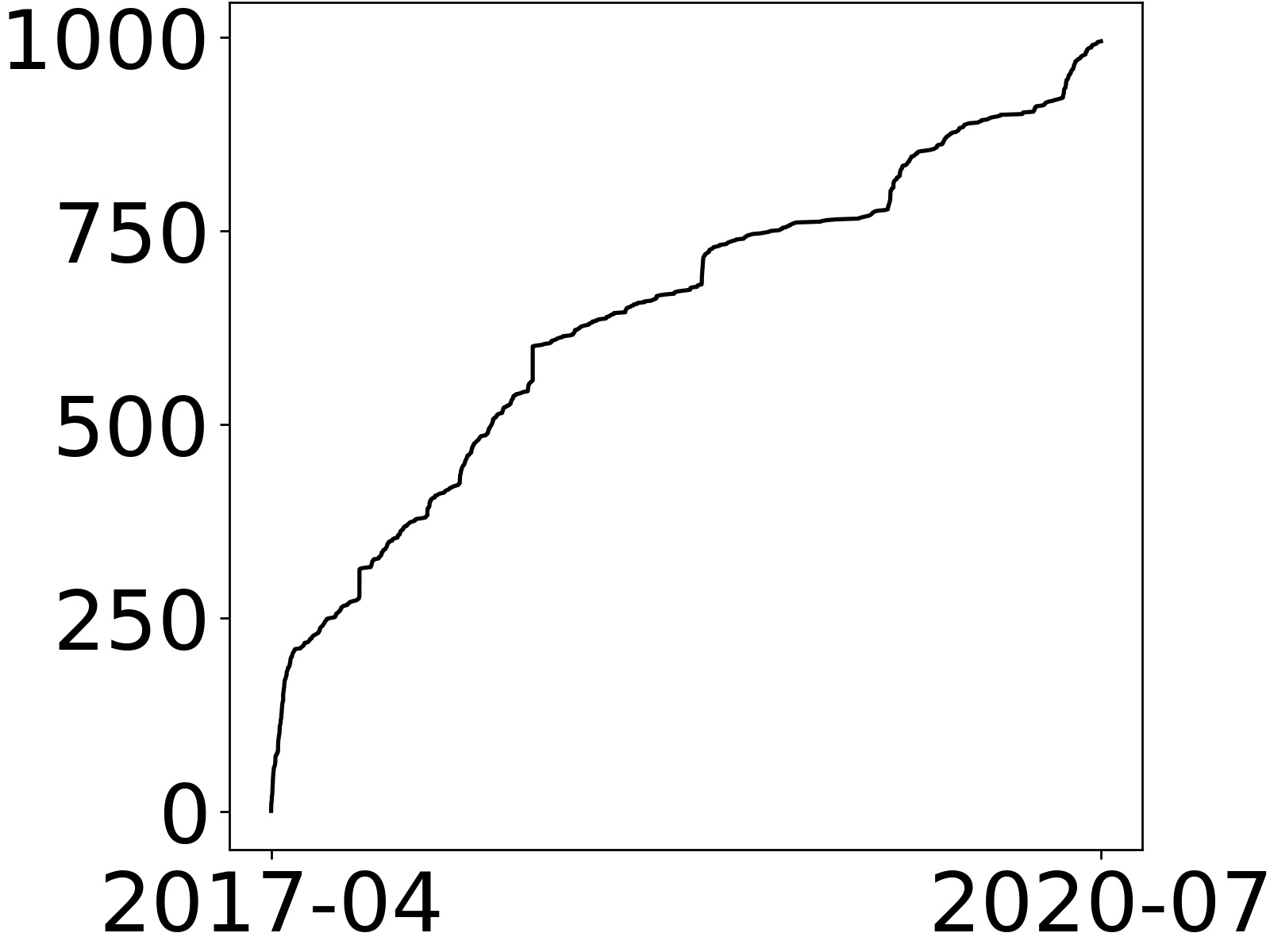}
		\minifigcaption{dlplibs}
	\end{subfigure}%
	\begin{subfigure}[b]{.166\linewidth}
		\centering
		\includegraphics[width=\linewidth]{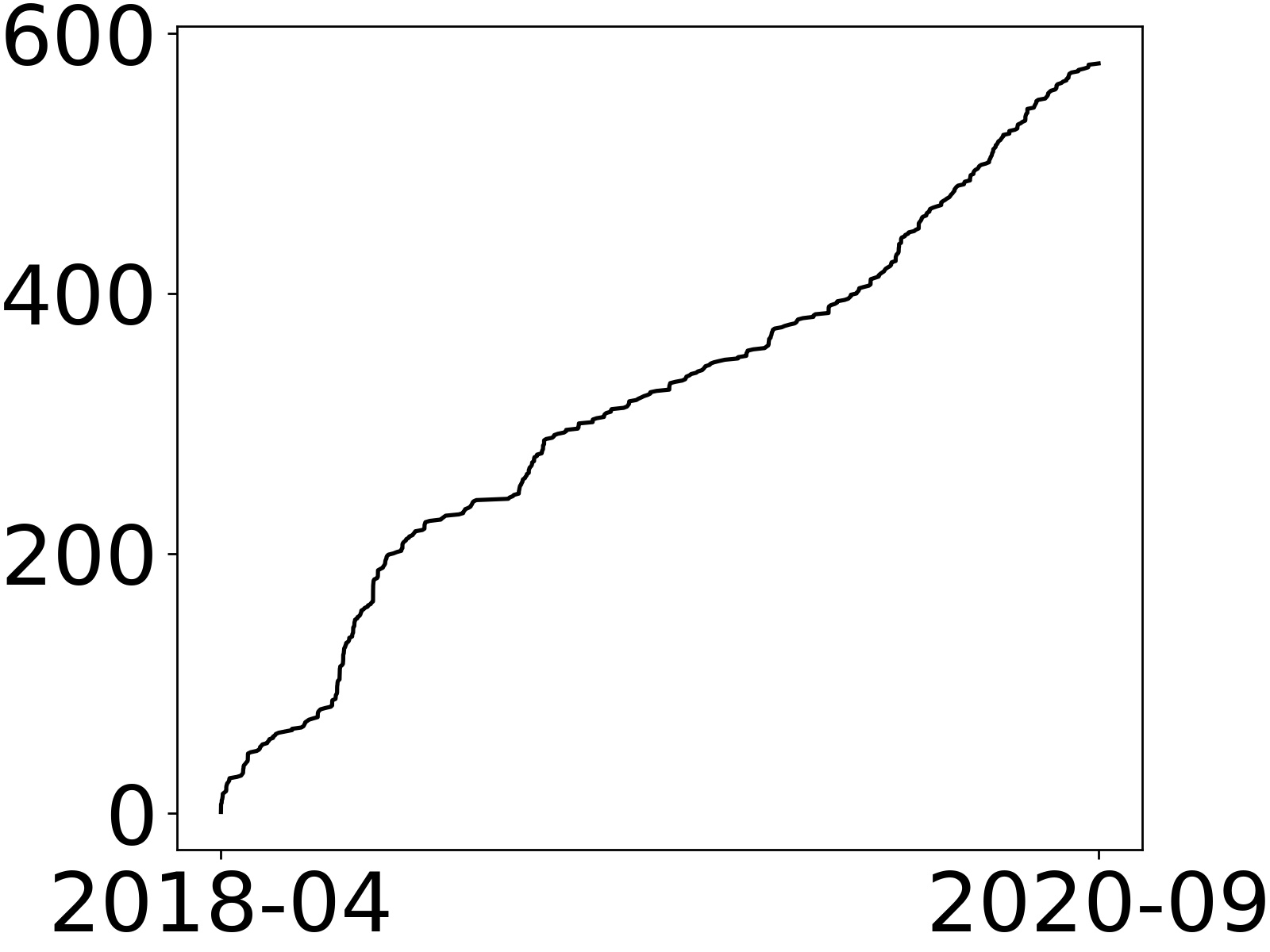}
		\minifigcaption{envoy}
	\end{subfigure}%
	\begin{subfigure}[b]{.166\linewidth}
		\centering
		\includegraphics[width=\linewidth]{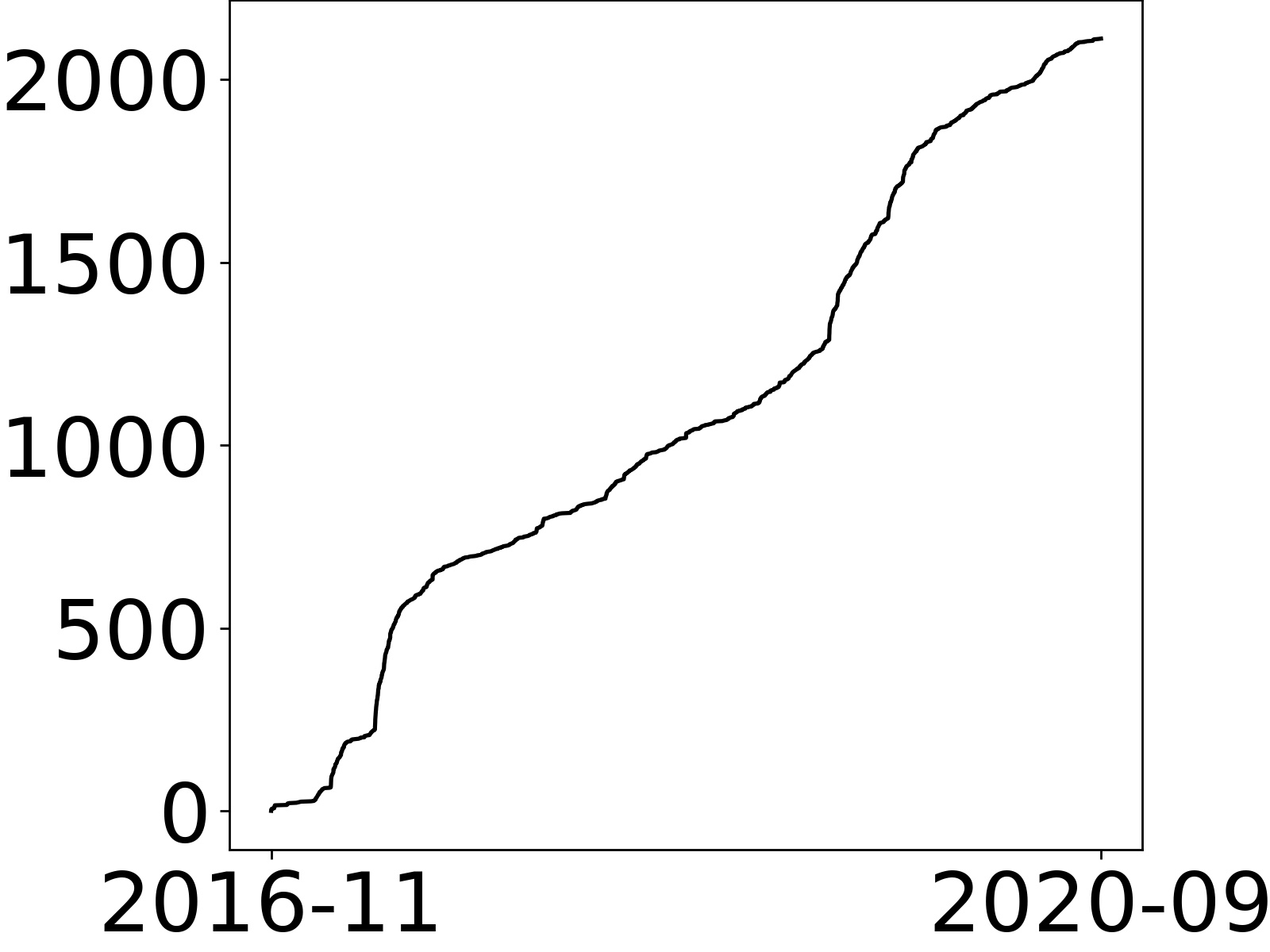}
		\minifigcaption{ffmpeg}
	\end{subfigure}%
	\begin{subfigure}[b]{.166\linewidth}
		\centering
		\includegraphics[width=\linewidth]{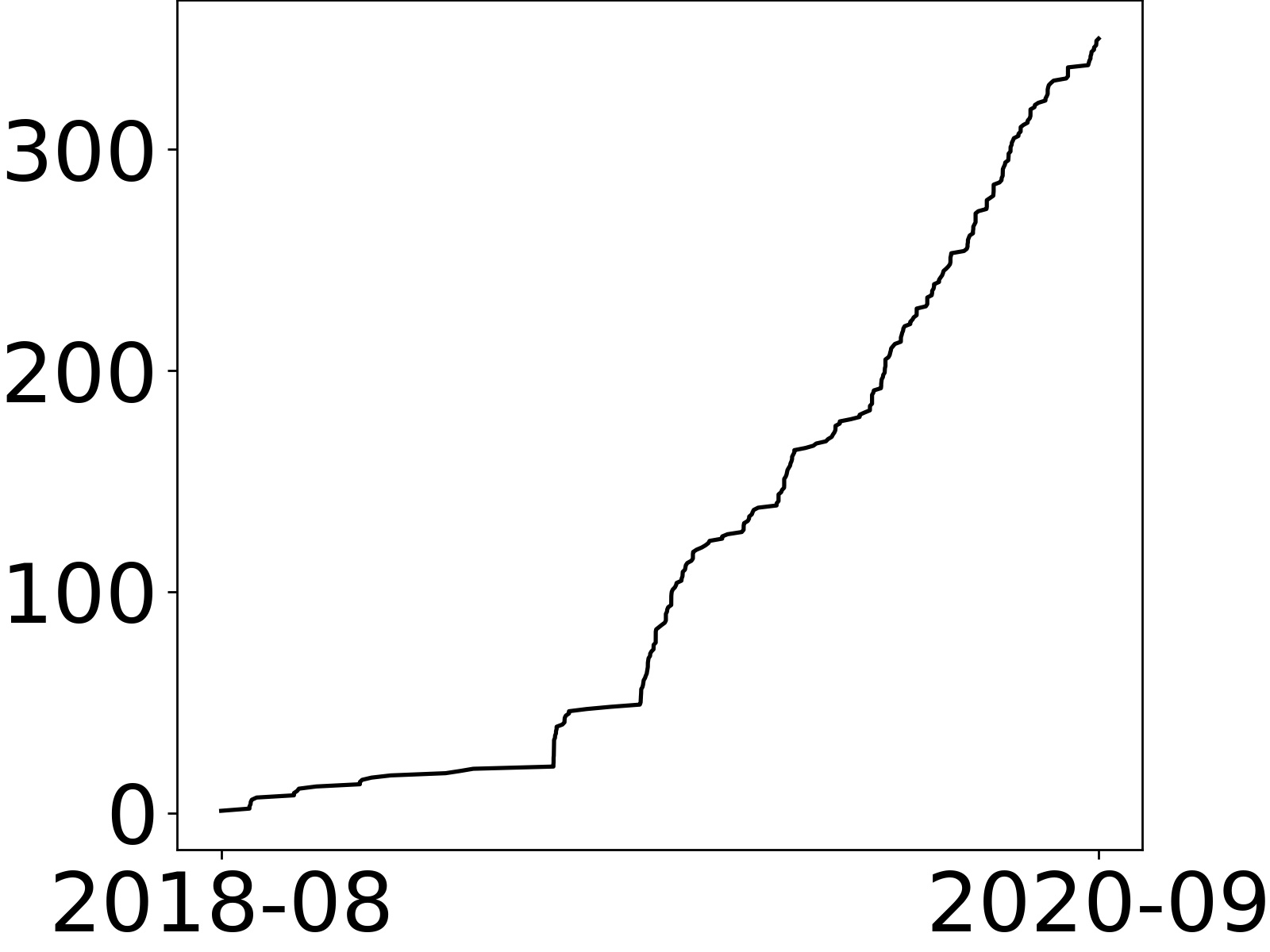}
		\minifigcaption{firefox}
	\end{subfigure}%

	\begin{subfigure}[b]{.166\linewidth}
		\centering
		\includegraphics[width=\linewidth]{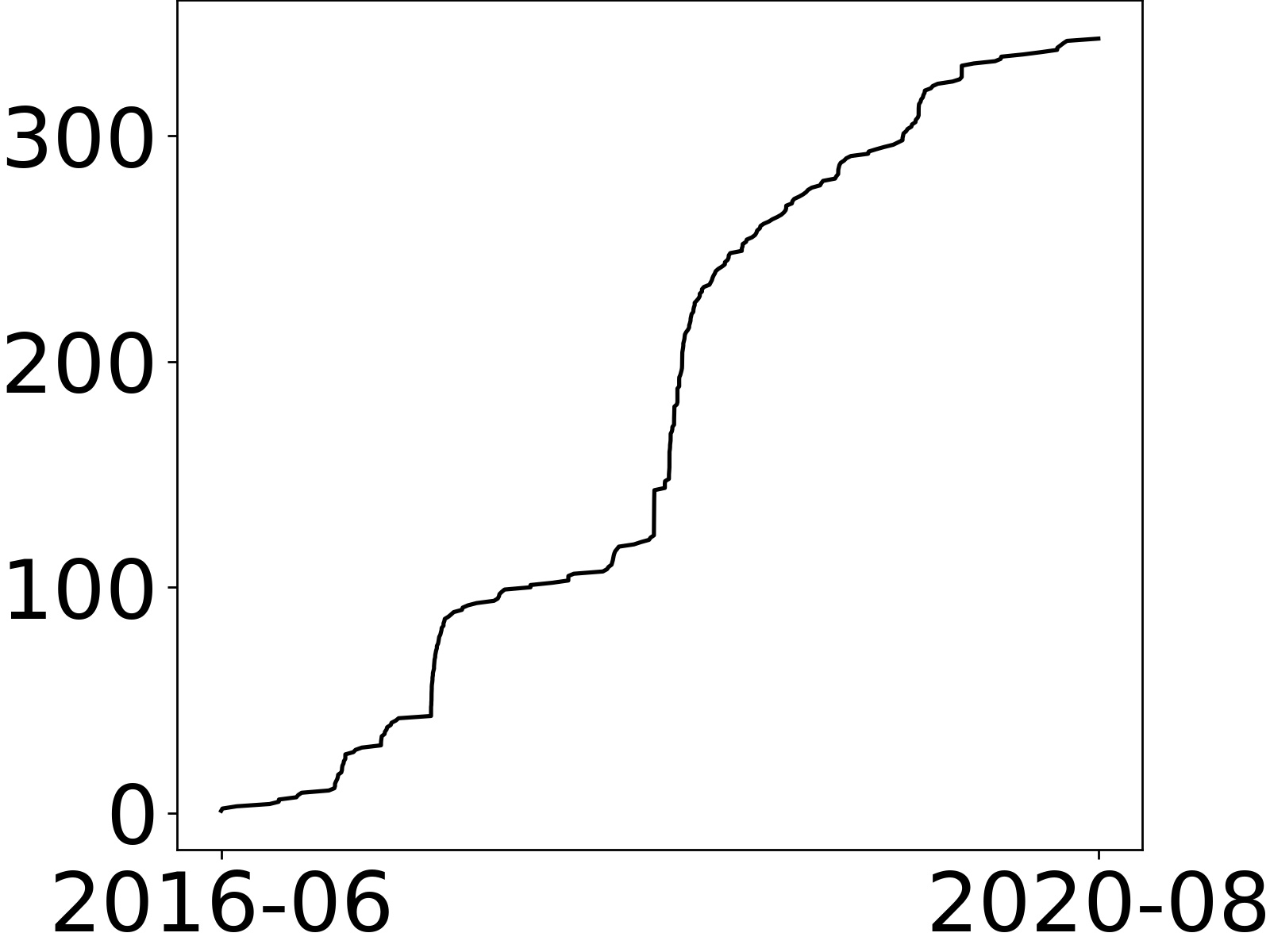}
		\minifigcaption{freetype2}
	\end{subfigure}%
	\begin{subfigure}[b]{.166\linewidth}
		\centering
		\includegraphics[width=\linewidth]{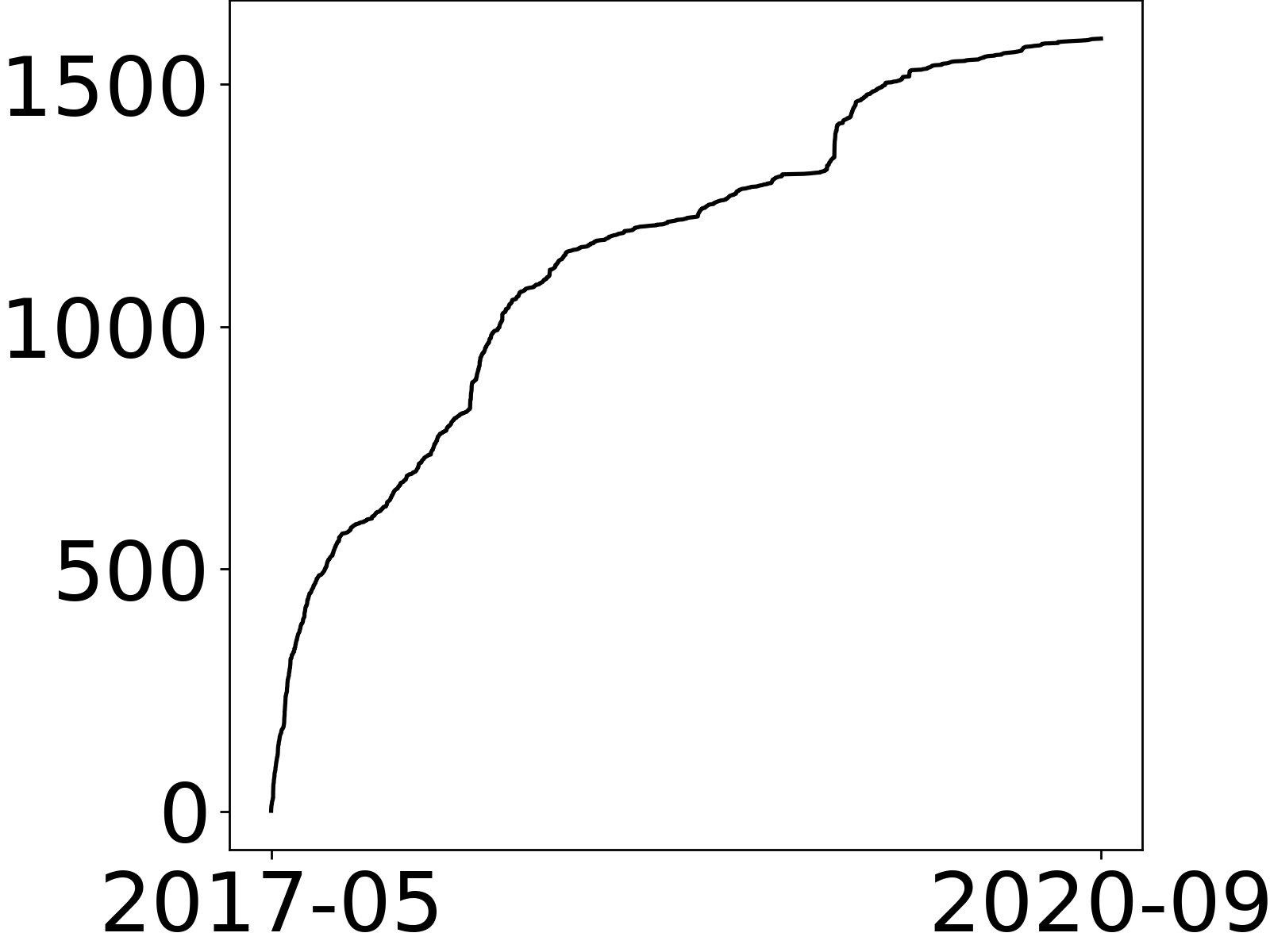}
		\minifigcaption{gdal}
	\end{subfigure}%
	\begin{subfigure}[b]{.166\linewidth}
		\centering
		\includegraphics[width=\linewidth]{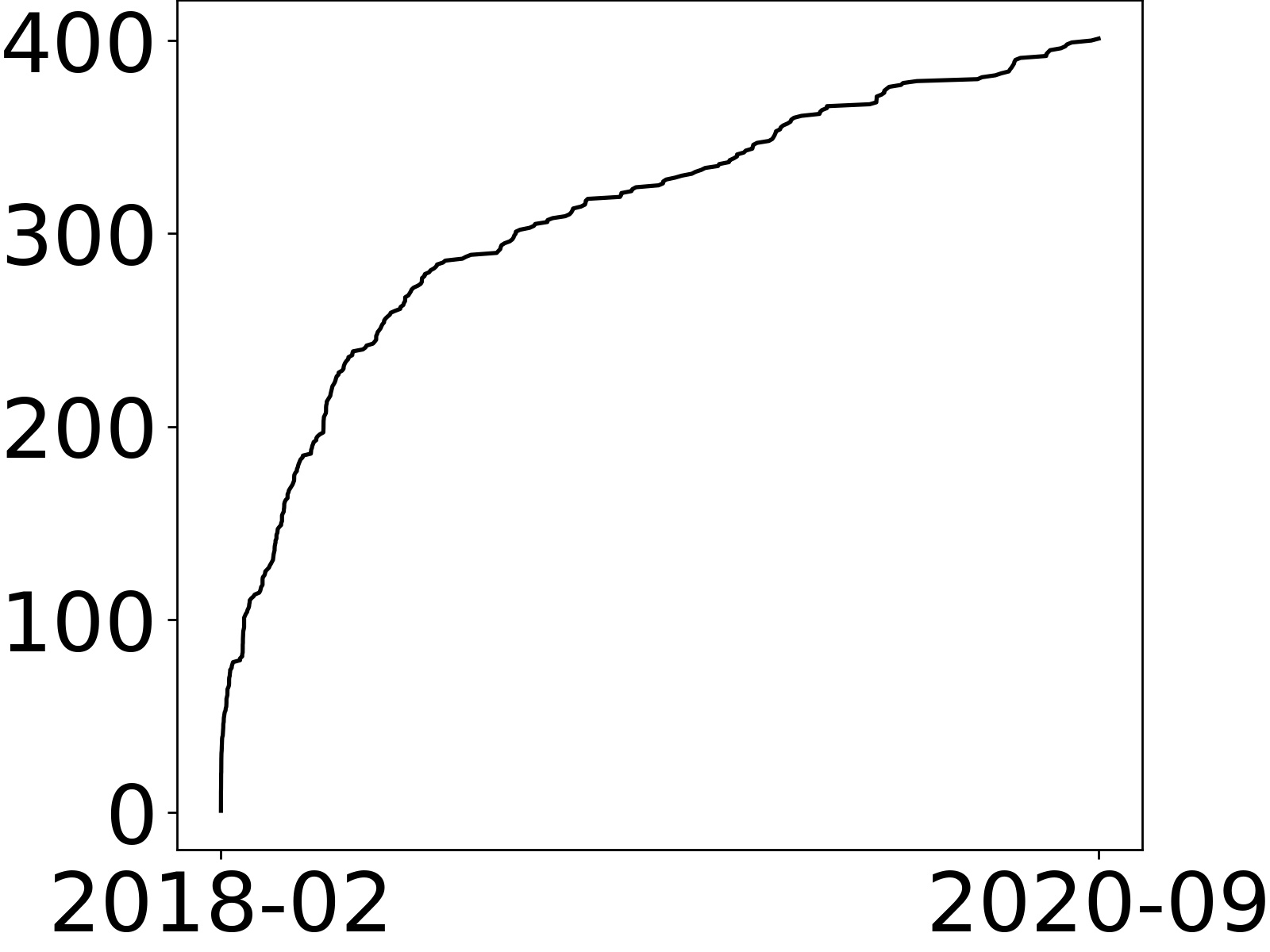}
		\minifigcaption{graphicsmagick}
	\end{subfigure}%
	\begin{subfigure}[b]{.166\linewidth}
		\centering
		\includegraphics[width=\linewidth]{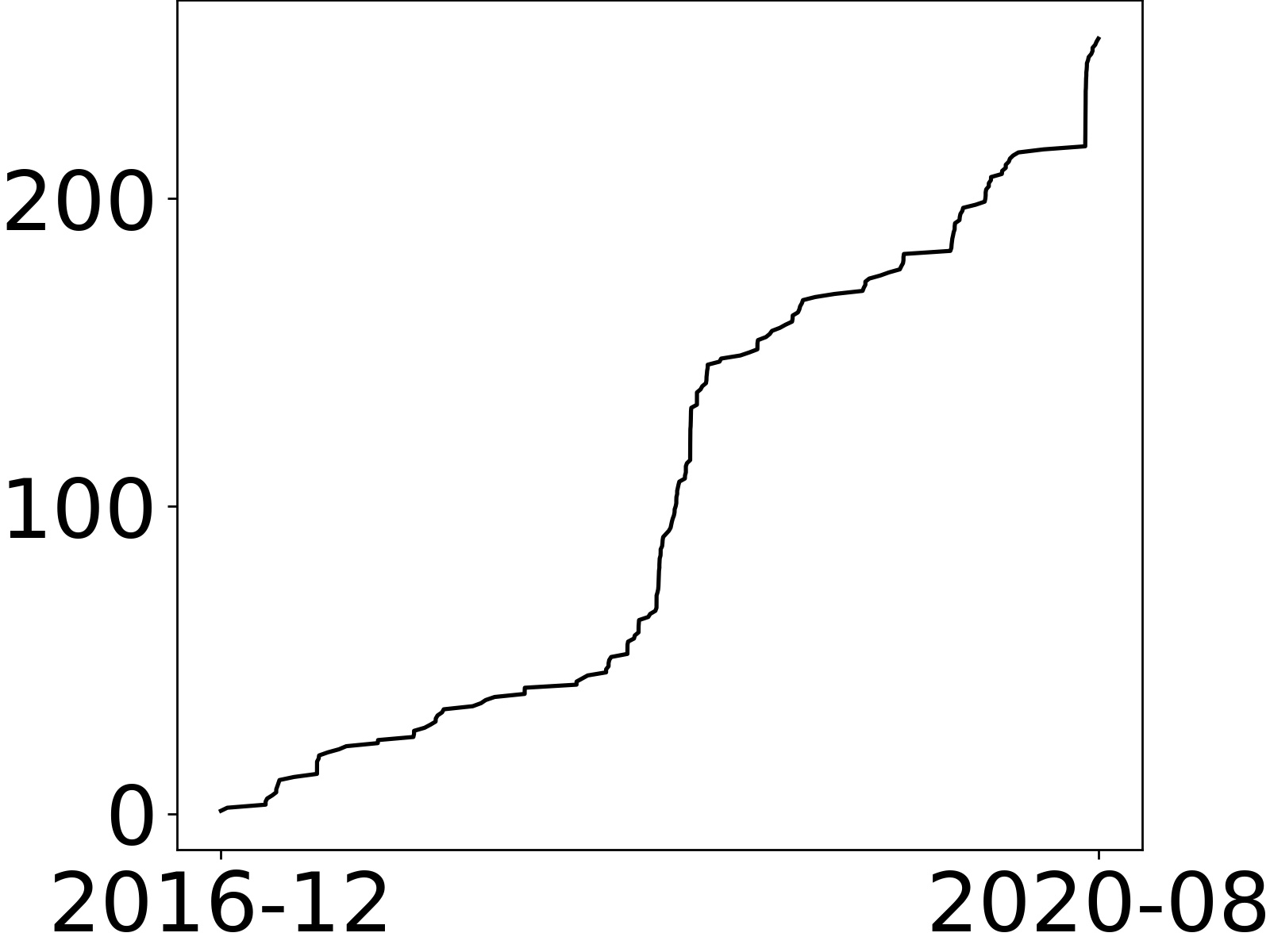}
		\minifigcaption{harfbuzz}
	\end{subfigure}%
	\begin{subfigure}[b]{.166\linewidth}
		\centering
		\includegraphics[width=\linewidth]{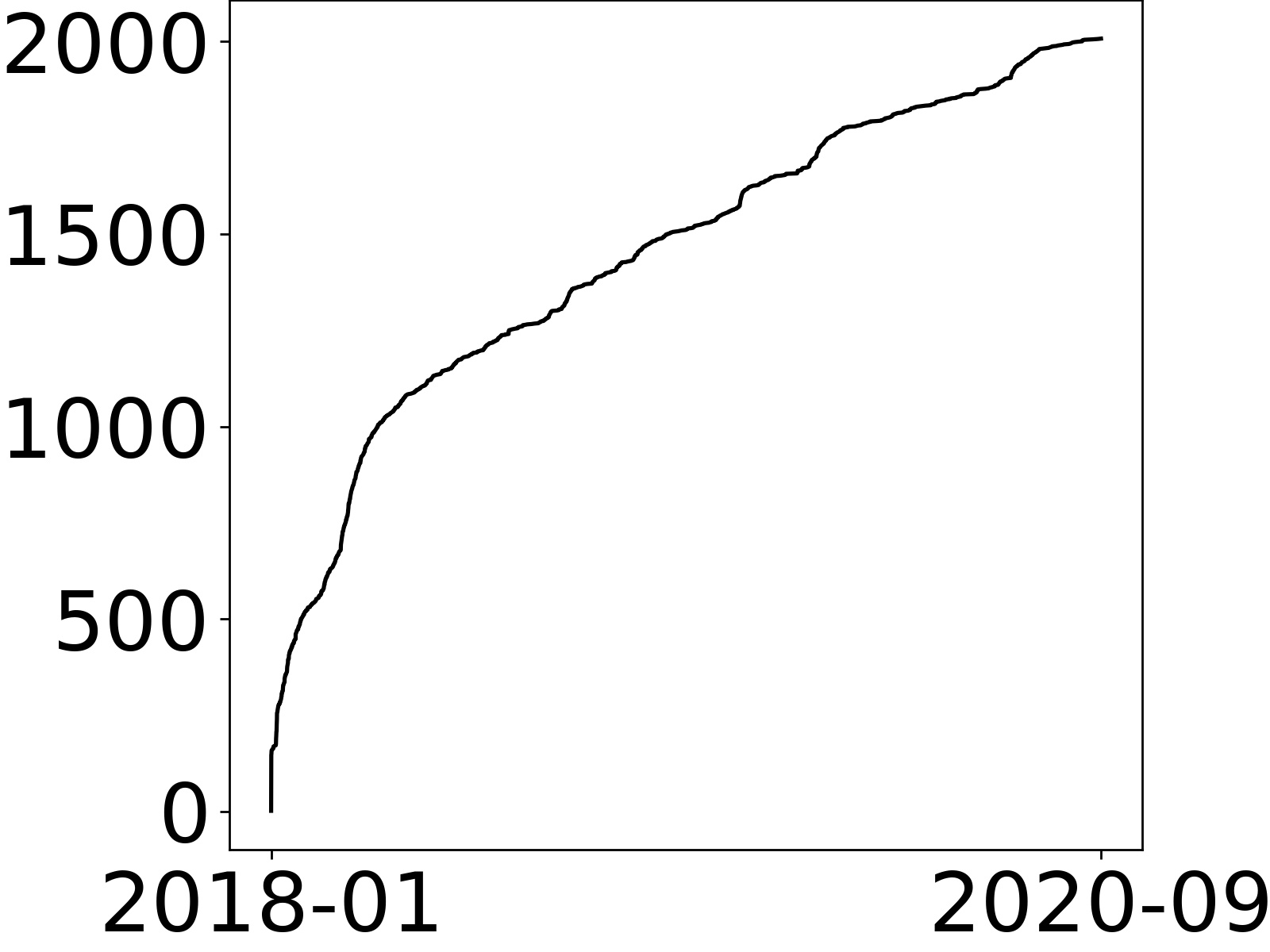}
		\minifigcaption{imagemagick}
	\end{subfigure}%
	\begin{subfigure}[b]{.166\linewidth}
		\centering
		\includegraphics[width=\linewidth]{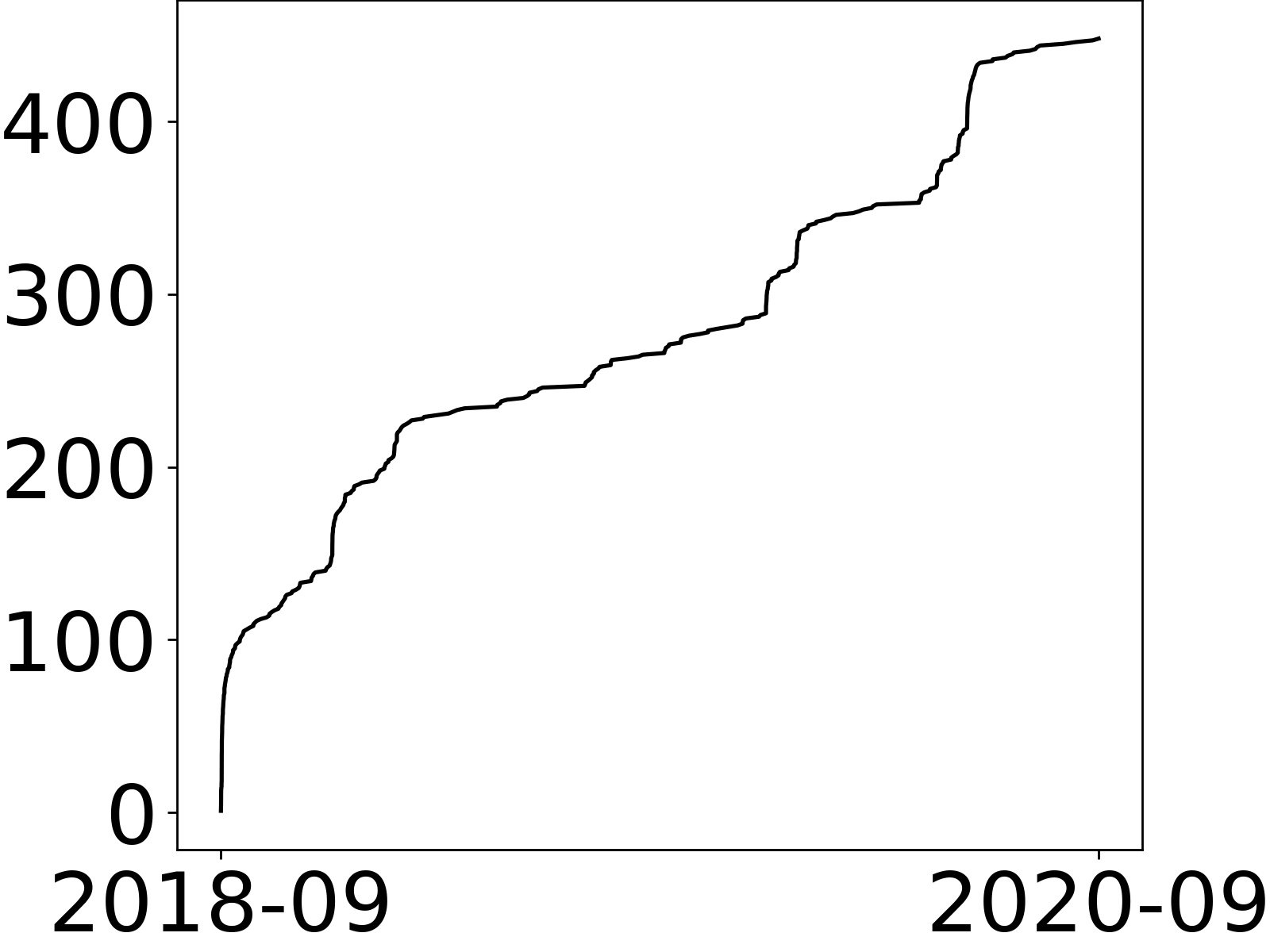}
		\minifigcaption{keystone}
	\end{subfigure}%

	\begin{subfigure}[b]{.166\linewidth}
		\centering
		\includegraphics[width=\linewidth]{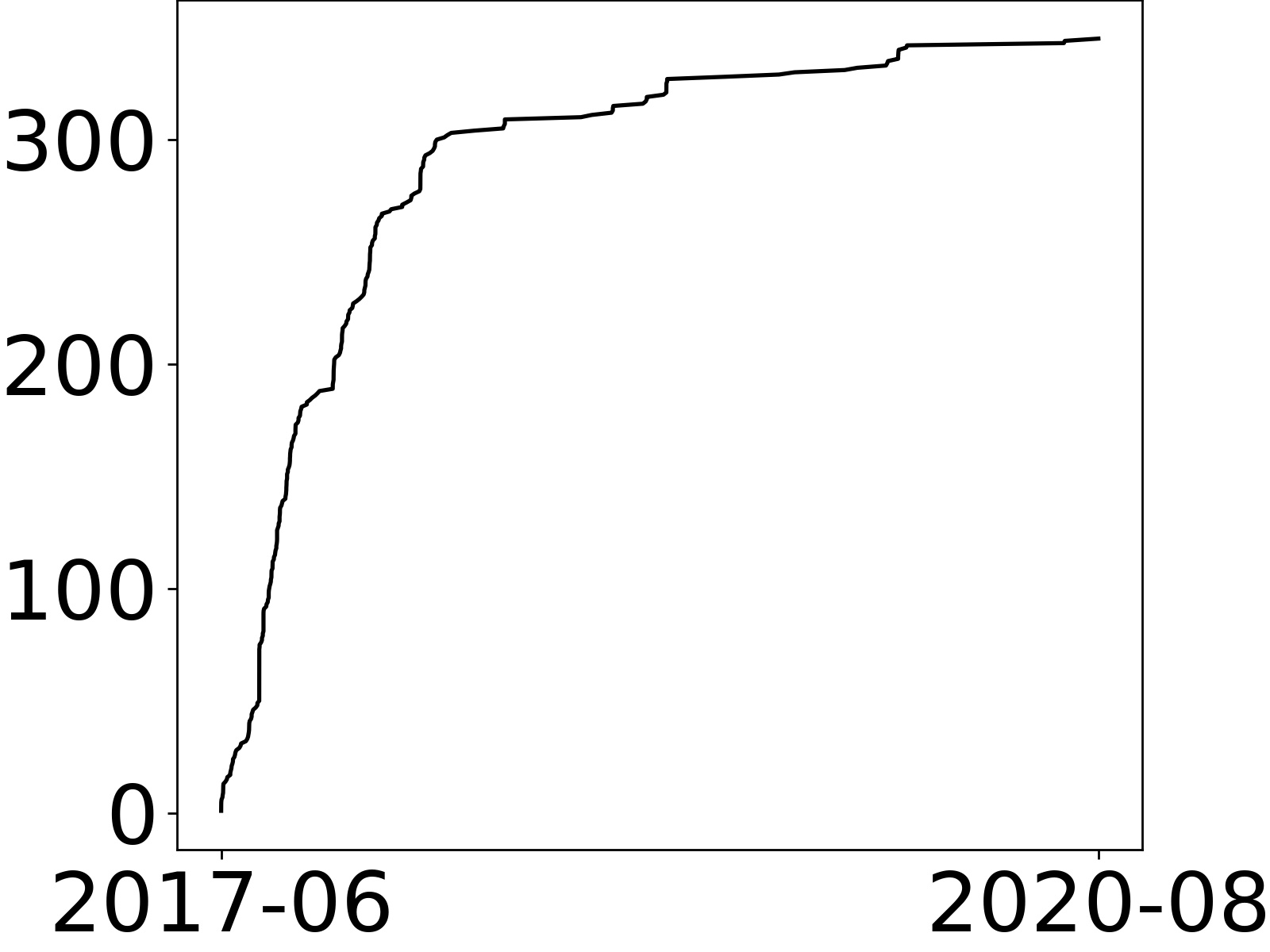}
		\minifigcaption{librawspeed}
	\end{subfigure}%
	\begin{subfigure}[b]{.166\linewidth}
		\centering
		\includegraphics[width=\linewidth]{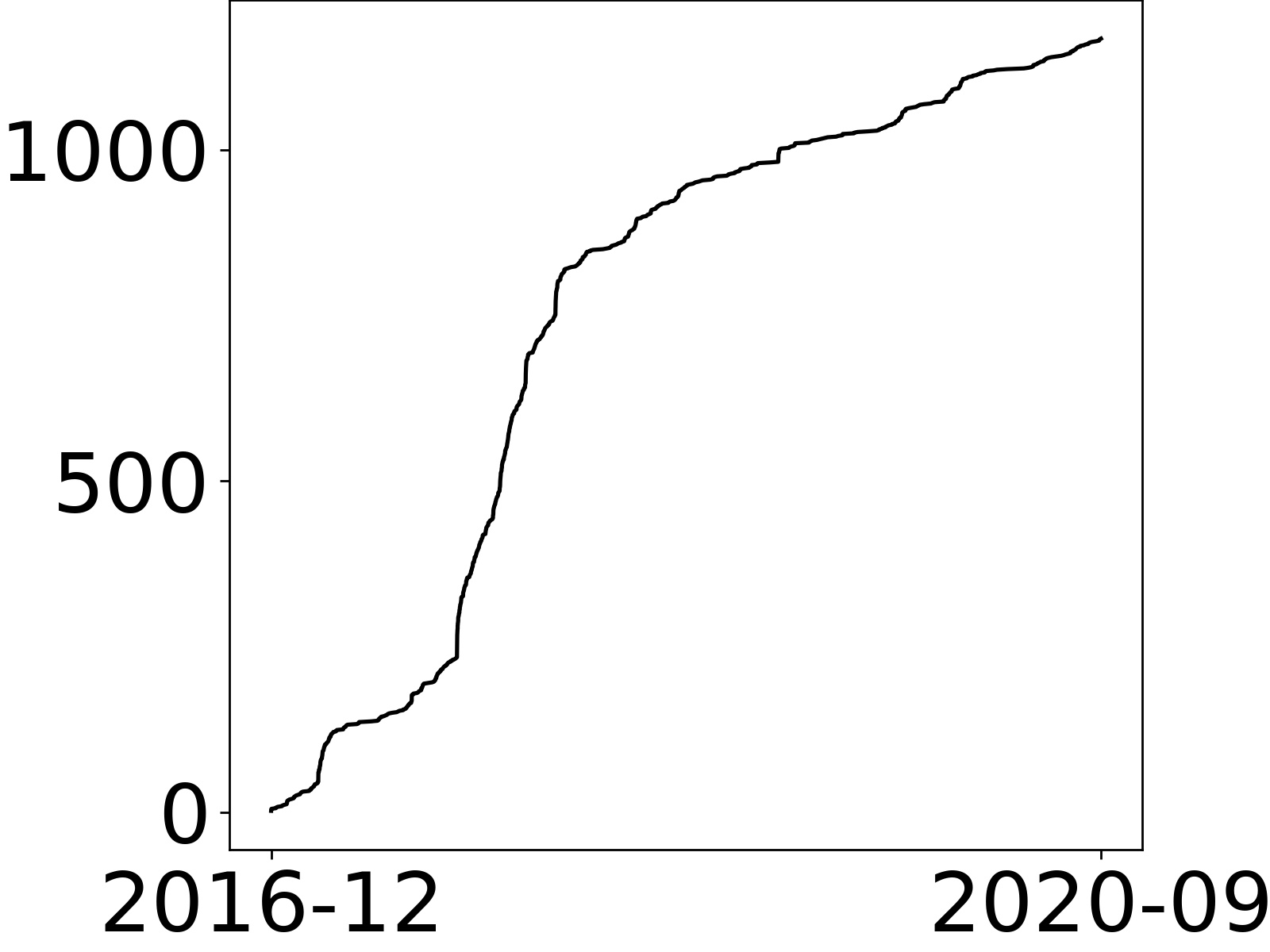}
		\minifigcaption{libreoffice}
	\end{subfigure}%
	\begin{subfigure}[b]{.166\linewidth}
		\centering
		\includegraphics[width=\linewidth]{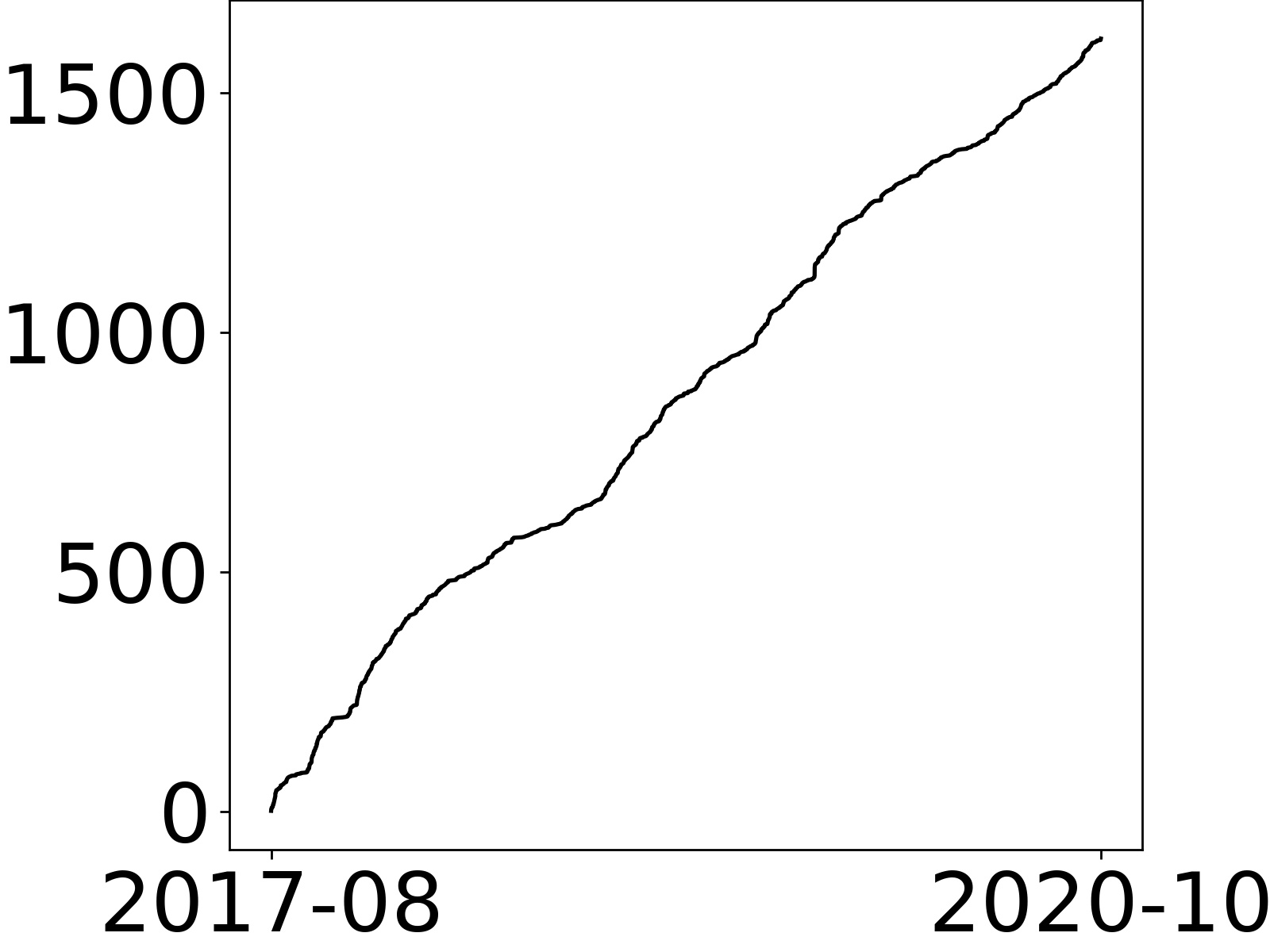}
		\minifigcaption{llvm}
	\end{subfigure}%
	\begin{subfigure}[b]{.166\linewidth}
		\centering
		\includegraphics[width=\linewidth]{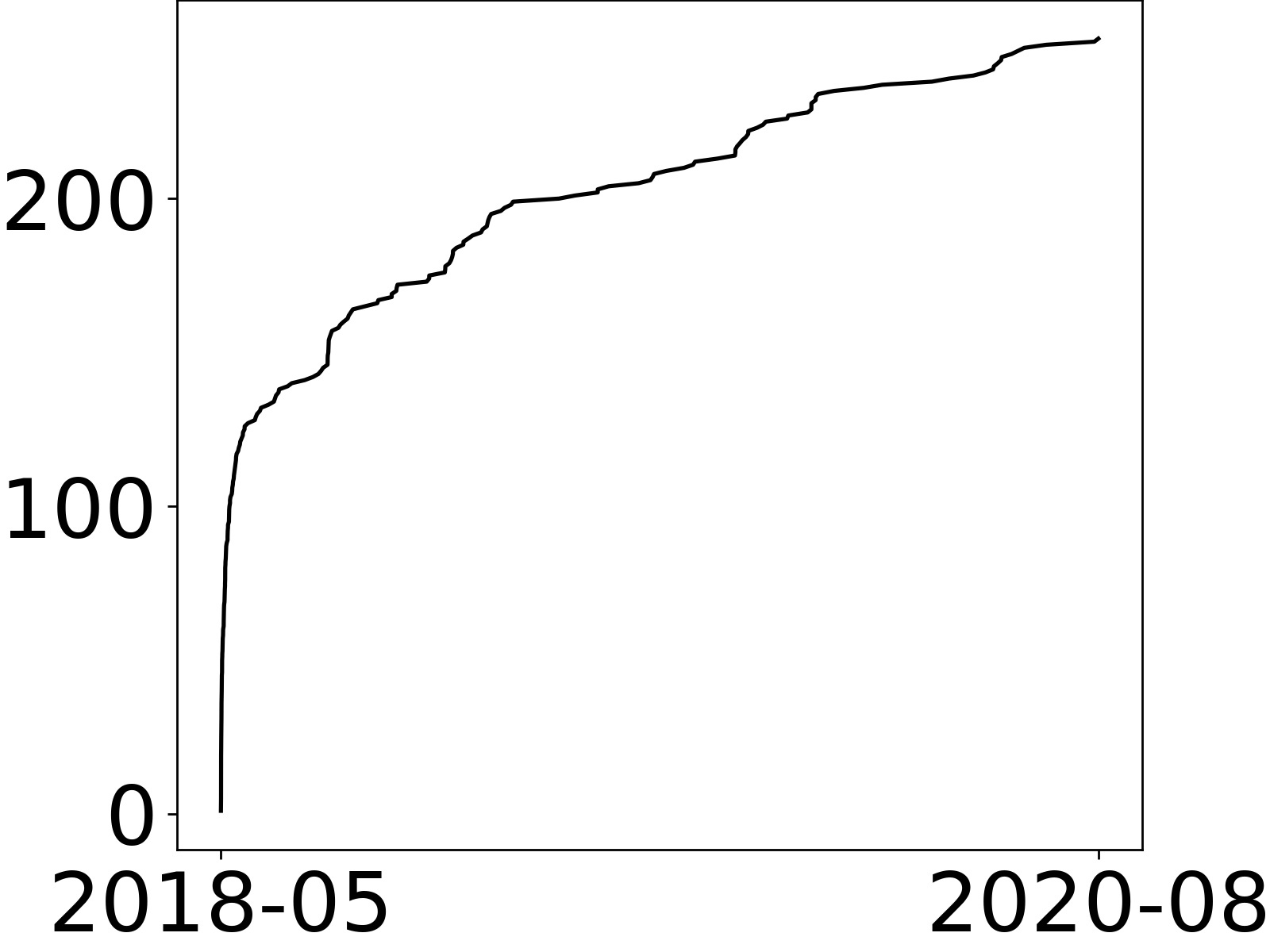}
		\minifigcaption{poppler}
	\end{subfigure}%
	\begin{subfigure}[b]{.166\linewidth}
		\centering
		\includegraphics[width=\linewidth]{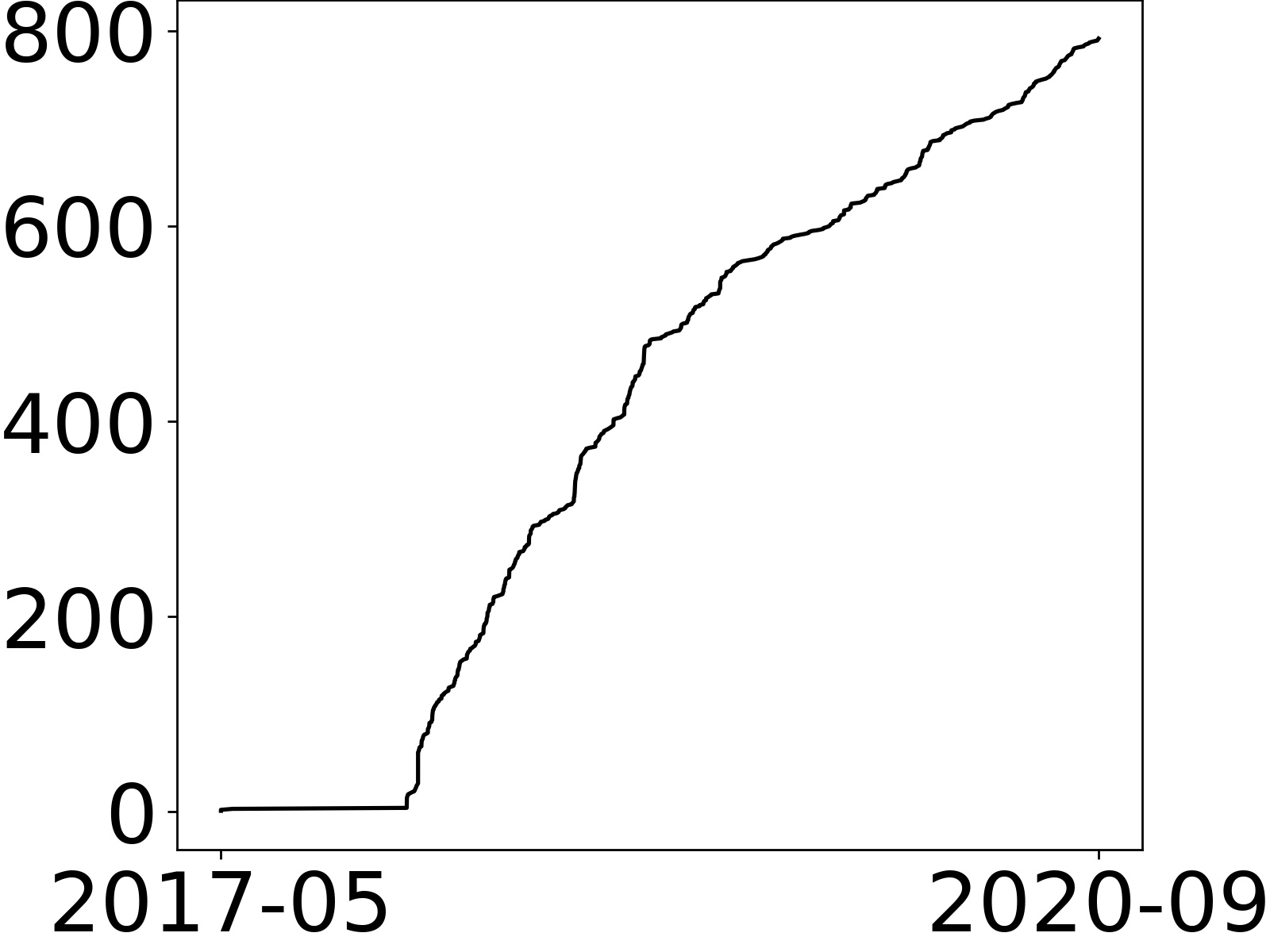}
		\minifigcaption{skia}
	\end{subfigure}%
	\begin{subfigure}[b]{.166\linewidth}
		\centering
		\includegraphics[width=\linewidth]{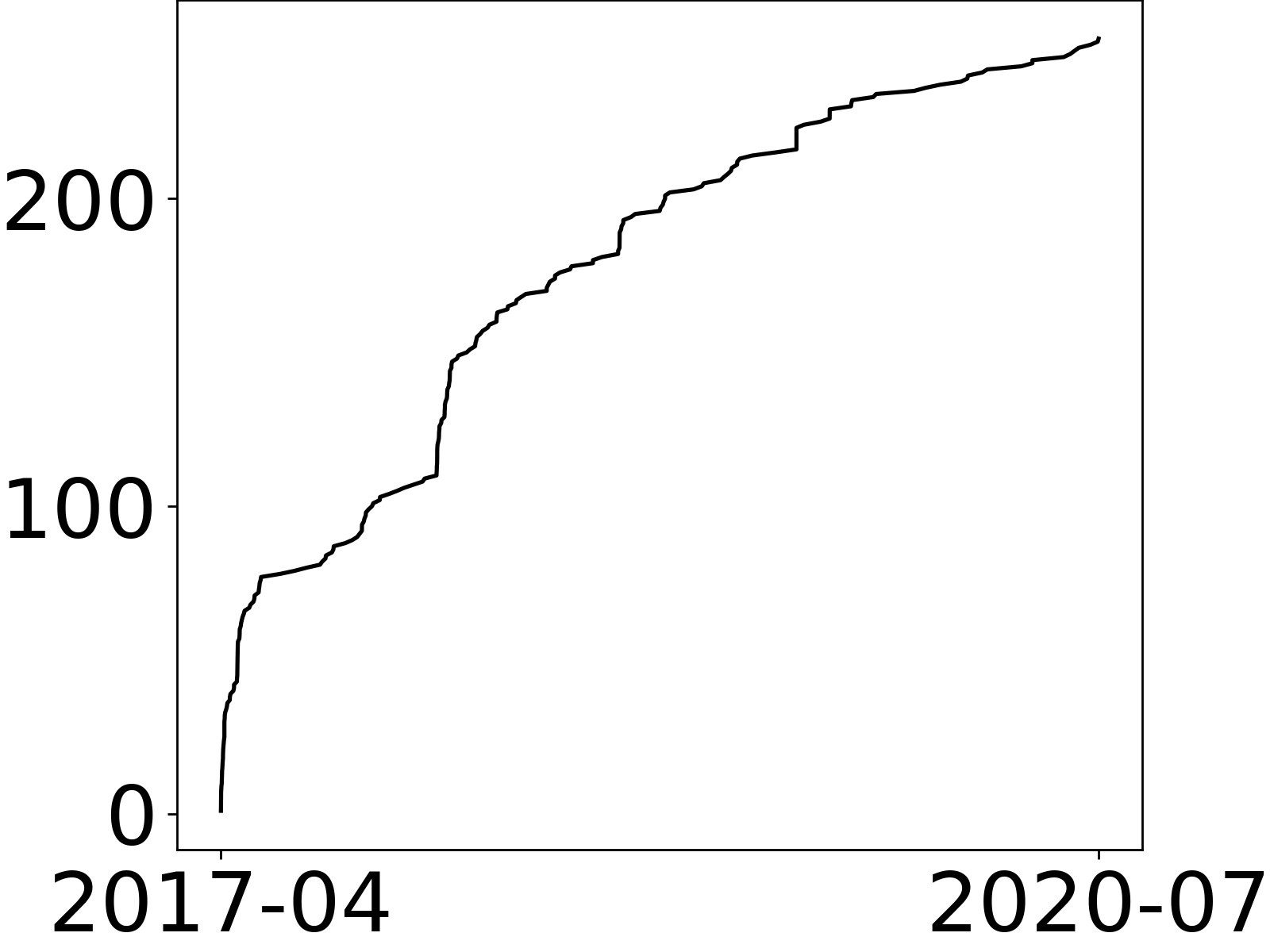}
		\minifigcaption{wireshark}
	\end{subfigure}%

	\captionsetup{aboveskip=-4pt} 
	\caption{Cumulative numbers of bugs found in projects with $>$200 bugs.
	The numbers of bugs found over time often exhibit punctuated equilibria, a common phenomenon
	in genetic algorithms such as those used in coverage-guided fuzzers.}
	\label{fig:bugs-found}
\end{figure*}

A motivation behind continuous fuzzing,
as opposed to one-off fuzzing campaigns, is the benefit of continuously finding new bugs
as software evolves. The continuous nature of OSS-Fuzz motivates
the following research question:

\textbf{RQ-L:} How does the bug discovery rate change over time?

\paragraph{Results} Figure~\ref{fig:bugs-found} show cumulative
numbers of bugs found in projects with $>$200 discovered
bugs. Many projects' cumulative distributions exhibit
\emph{punctuated equilibria}, with periods of slow growth punctuated by bursts
of rapid growth.

Punctuated equilibria~\cite{gould-punctuated-equilibria}
appear in genetic algorithms~\cite{vose-punctuated-equilibria}
--- which AFL, libFuzzer, and honggfuzz use as part of their search strategy
--- and in software evolution~\cite{gorshenev-software-evolution, wu-thesis, icse01-anton}.
Fuzzers also exhibit punctuated equilibria in the
long-term growth in coverage~\cite{mtfuzz, collafl, neuzz, chornozon, angora, pldi17-glade, kAFL, redqueen, csifuzz,
esecfse20-boehme, crfuzz, matryoshka, greyone, zeller-fuzzing-book}
and number of bugs found~\cite{collafl, rsda19-kernel-fuzz, ccs18-eval-fuzz, usenixsec12-fuzz-frags, angora, kAFL,
redqueen, greyone}.
We present evidence of punctuated equilibria in multi-year continuous
fuzzing campaigns.

ClusterFuzz prioritizes hardware resources towards fuzz targets that
are actively discovering new coverage~\cite{blackhat19-clusterfuzz}.
Such a selection strategy increases selection pressure, which steepens peaks
and flattens plateaus in the number of bugs found.

\takeaway{Bug discovery over time often exhibit punctuated equilibria, with
short bursts of rapid bug discovery, rather than a consistent trickle of bugs.}

\section{Discussion and Implications}
\label{sec:discussion}

We are encouraged to see that OSS-Fuzz quickly finds regressions and developers
quickly fix them. Our results provide real-world evidence of continuous fuzzing's
effectiveness.

Flaky bugs, however, are problematic for developers, even if the bug appears
often. The most flaky fault types, timeout and out of memory,
may owe their flakiness from unpredictable resource availability or usage.
Flakiness is also a symptom of a failure to control non-deterministic behavior
during fuzzing, and flaky bugs may point developers to non-deterministic
code that might require remediation for effective testing.

Timeouts and out of memory errors, respectively the first and third most common
fault types, stand out as problematic. Even among non-flaky bugs,
timeouts and OOMs are slower to detect, more often unfixed, and slower to fix.
Specialized fuzzers~\cite{slowfuzz,perffuzz,fuzzfactory} can
find more such bugs in less time (countering the slow time-to-detect);
however, existing widely-used fuzzers are already finding many such bugs.
The lower severity of resource exhaustion bugs, combined with the potential
human annoyance of reproducing such bugs, likely contributes to developers
ascribing lower priority to such bugs.

Memory leaks and assertion violations, two other low-severity fault types,
are common in fuzz blockers, which block fuzzing downstream from the bug.
To mitigate the blockage, a continuous fuzzing system can temporarily disable
leak detection or assertions to fuzz past the blockers.
Developers fix fuzz blockers more slowly, and these automatic mitigations offer
stopgap solutions until someone fixes the blockers.

On the other end of the severity scale, despite OSS-Fuzz finding thousands
of severe bugs, such as memory corruption vulnerabilities, very few bugs received CVEs.
This is a weakness in open-source security, as tools that rely on CVEs as
a source of threat intelligence are not aware of potential threats in open-source
software, which can percolate to computer systems and other software that depend on compromised
open-source software. Developers without a security background might not be aware
of the impact of CVEs, or might not wish to navigate through the process of requesting
CVEs. More awareness and guidance may help to alleviate the issue.

Since fuzzing campaigns often exhibit punctuated equilibria with bursts of rapid
bug discovery, developers may get an unpleasant surprise due to an avalanche of bugs
in a short timeframe. A rapid dump of bugs can overwhelm developers and elicit
defensive behavior~\cite{fuzzcon20-holler}. Alerting developers of this phenomenon
would mentally prepare them, avoiding surprises.

Fuzzing researchers expressed need for a fair time budget when evaluating fuzzers~\cite{shonan-fuzzing}.
We suggest five days, OSS-Fuzz's median time-to-detect, as an option.
Short time budgets (e.g., one hour) can introduce bias towards certain fault types.

\paragraph{Limitations} OSS-Fuzz is a continuous fuzzing service for open source software.
Our findings might not extend to one-off fuzzing, commercial
software, continuous fuzzing frameworks other than ClusterFuzz, or
fuzzers other than the coverage-guided fuzzers AFL, libFuzzer, and honggfuzz.

\section{Related Work}
\label{sec:related-work}

Our work joins the corpus of empirical studies of software bugs,
which have also examined fault types~\cite{li-sec-patches,liu-vuln-distribution,shahzad-vuln-lc,asid06-li,
    verdi-stack-overflow-vulns,msr12-perf-bugs,msr11-sec-vs-perf,cacm90-fuzzing},
flakiness or irreproducibility~\cite{msr14-works4me,fse14-flaky-tests,icse20-flaky-lifecycle,
    usenixsec18-vuln-reproducibility,msr12-perf-bugs},
unfixed bugs~\cite{icse10-which-gets-fixed,icsme18-which-gets-fixed,msr12-perf-bugs},
CVEs~\cite{li-sec-patches,liu-vuln-distribution,shahzad-vuln-lc,issre10-cve,lsad06-vulns},
and fault lifecycles~\cite{csmr14-long-lived-bugs,msr14-dormant-bugs,icse20-flaky-lifecycle,msr11-sec-vs-perf,
    msr12-perf-bugs,wcre12-time2fix,lsad06-vulns}.
In particular, Miller et al.~\cite{cacm90-fuzzing}, an empirical study of the reliability of Unix utilities,
coined the term ``fuzz'' to denote a tool for randomly generating test inputs; they examined
fault types of fuzzer-found bugs, similar to our empirical analysis of OSS-Fuzz.

MITRE's 2020 CWE Top 25~\cite{cwe-top-25} lists the most impactful common
software weaknesses observed in 2018--2020. Out-of-bounds write (CWE-787),
out-of-bounds read (CWE-125), improper restrictions within memory buffer bounds (CWE-119),
use after free (CWE-416), integer overflow (CWE-190),
null dereference (CWE-476), and uncontrolled resource consumption (CWE-400)
rank among the top 25. OSS-Fuzz found all of the above.

The 2019 Shonan Meeting on Fuzzing and Symbolic Execution~\cite{shonan-fuzzing}
identified a need for more empirical analysis on fuzzing.
They expressed interest in difficult or ``deep'' bugs,
fair time budgets for evaluating fuzzers, and
human-fuzzer interaction.
Our empirical work sheds light on the current state of fuzzing in practice,
illuminating the baseline to improve on.

We grow the literature on how practitioners interact with fuzzing.
An industry report~\cite{saner18-liang} found that writing
fuzz targets required training, blockers are obstacles,
and dirty hacks (e.g., disabling error reporting) can hide bugs from fuzzers.

Prior work examined continuous fuzzing for OS kernels.
syzbot~\cite{syzbot} is a continuous fuzzing system for kernels.
A study~\cite{rsda19-kernel-fuzz} examined
2269 syzbot-found bugs in Linux, FreeBSD, NetBSD, and OpenBSD .
The study analyzed the time to fix bugs (median of 38 days for Linux,
$<$20 days for BSD) and fault types (debug checks, assertions,
and use after free were among the most common).
A report~\cite{esecfse19-shi} on implementing continuous fuzzing for enterprise
Linux kernels identified fuzz blockers as obstacles and found
132 bugs, 41 of which were reproducible. Lockups, deadlocks, and warnings
were common fault types.
We augment the existing literature on continuous fuzzing by analyzing
a larger dataset of bugs in a diverse set of open source software.

\section{Conclusion}
\label{sec:conclusion}

We conduct the first empirical analysis of OSS-Fuzz bugs, evaluating 23,907
fuzzer-found bugs spanning over 4 years and 316 software projects.
We fill a need for empirical evaluations of fuzzing,
shine light on the state of the practice,
and unveil insights for research and practice.
While we examine bug reports, OSS-Fuzz and its participant projects
contain other artifacts, such as fuzz targets, configurations,
and code commits. We suggest an examination of these artifacts as part of a
deeper study of continuous fuzzing. We provide
an open-science package: \url{https://doi.org/10.5281/zenodo.4625207}

\section*{Acknowledgements}
We thank the ClusterFuzz and OSS-Fuzz teams, plus their OSS collaborators,
for improving the quality of open source.
We thank Rohan Padhye for his valuable feedback and insights.
This research was partially funded by AFRL (\#OSR-4066) and the NSF (\#CCF-1750116).
The authors are grateful for their support. Any opinions or findings expressed
are those of the authors and do not necessarily reflect those of the US Government.

\bibliographystyle{IEEEtran}
\bibliography{references}

\end{document}